\documentclass[10pt, journal, final]{IEEEtran} 
\pdfoutput=1
\IEEEoverridecommandlockouts

\usepackage{amsmath,bm}
\usepackage{amsfonts}
\usepackage{graphicx}
\usepackage{epstopdf}
\usepackage{color}
\usepackage{subeqnarray}
\usepackage{algorithmic}
\usepackage{algorithm}
\usepackage{indentfirst} 
\usepackage{cite}
\usepackage{tikz}
\usepackage{multirow}
\usepackage{mathtools}
\usepackage{hyperref}
\usepackage{multirow}
\newcommand{\review}[1]{\textcolor{black}{#1}}

\newcommand{\software}[1]{{\tt#1}}
\newcommand{\norm}[1]{\lVert#1\rVert}

\begin{document}
\title{A Learning-based Nonlinear Model Predictive Controller for a Real Go-Kart based on Black-box Dynamics Modeling through Gaussian Processes}
\author{Enrico~Picotti$^{1}$,~\IEEEmembership{Graduate~Student~Member,~IEEE}, Enrico~Mion$^{2}$, Alberto~Dalla~Libera$^{1}$, Josip~Pavlovic$^{2}$, Andrea~Censi$^{2}$, Emilio~Frazzoli$^{2}$,~\IEEEmembership{Fellow,~IEEE}, Alessandro~Beghi$^{1}$,~\IEEEmembership{Senior~Member,~IEEE}, Mattia~Bruschetta$^{1}$
\thanks{$^{1}$Department of Information Engineering, Università di Padova, Italy \newline {\tt\small \{name.surname\}@dei.unipd.it}
\newline$^{2}$Institute for Dynamic Systems and Control, ETH Zürich,
Switzerland, {\tt\small \{name.surname\}@ethz.ch}
\newline This research involved a single driver on the go-kart in order to gather the needed data and test the controller. 
The gathered data have been anonymized and processed upon informed consent of the user about the purposes and methods of the data collection.}}

\maketitle

\begin{abstract}
Lately, Nonlinear Model Predictive Control (NMPC) has been successfully applied to (semi-) autonomous driving problems and has proven to be a very promising technique. However, accurate control models for real vehicles could require costly and time-demanding specific measurements.
To address this problem, the exploitation of system data to complement or derive the prediction model of the NMPC has been explored, employing \textit{learning dynamics} approaches within Learning-based NMPC (LbNMPC).
Its application to the automotive field has focused on discrete grey-box modeling, in which a nominal dynamics model is enhanced by the data-driven component.
In this manuscript, we present an LbNMPC controller for a real go-kart based on a continuous black-box model of the accelerations obtained by Gaussian Processes. We show the effectiveness of the proposed approach by testing the controller on a real go-kart vehicle, highlighting the approximation steps required to get an exploitable GP model  on a real-time application.

\end{abstract}

\section{Introduction}

In recent decades, the development of sophisticated advanced driver assistance systems and the increased use of (semi-) autonomous driving have led to growing interest in four-wheel vehicle controllers, in both industrial and academic research.

In driving tasks, Model Predictive Control (MPC) strategy has been successfully applied, such as for path following and vehicle control \cite{Liniger2014, verschueren2016time, frasch2013auto, carvalho2013predictive, brus2019driver, picotti2022nonlinear}. MPC is an advanced control technique that, based on a plant model and constraints, optimizes the performance of the closed-loop system. 
However, plant characterization plays a crucial role in the optimization, and several models have been proposed for 4-wheels vehicles, ranging from complex dynamics models to simple bicycle models. Each formulation has advantages and disadvantages in terms of computational resources and prediction capabilities. Moreover, a precise characterization of car dynamics involves proper identification of tires, drag and friction forces, suspensions, steering systems, etc., which may require expensive and time-consuming specific measurements. 
Therefore, the exploitation of IMU and localization data to infer the vehicle dynamics could remove the necessity of these explicit analyses.
To cope with this modeling issue, \textit{learning dynamics} approaches can be used within the Learning-based Nonlinear Model Predictive Control (LbNMPC) framework, i.e. the combination of data-driven techniques and Nonlinear Model Predictive Control (NMPC) strategy \cite{hewing2020learning}.

\begin{figure}[t]
    \centering
    \vspace{0.1cm}
    \includegraphics[width=0.6\linewidth]{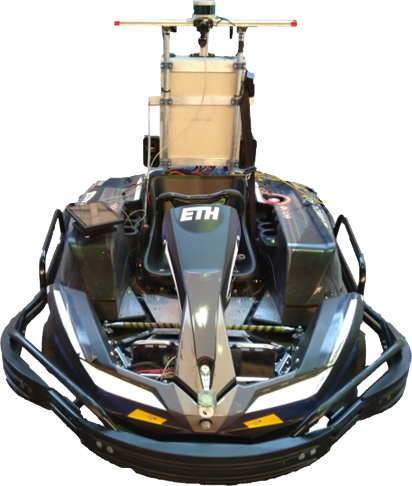}
    \caption{The go-kart platform.}
    \label{fig:gokart}
\end{figure}

The methods of \textit{learning dynamics} focus on the prediction model of the NMPC, which can be fully derived or partially improved through the exploitation of system data, leading to black-box or grey-box modeling, respectively \cite{aswani2013provably,kabzan2019learning,sforni2021learning,Miklos2021grey,carron2019data,piche2000nonlinear, MANZANO2020Robust, Coulson2019Data, maddalena21kpc,ostafew2014learning}. 
In this context, Gaussian Process Regression (GPR) has proven to be remarkably effective thanks to offline and runtime reduction techniques that maintain the GP characteristics, e.g. continuity and smoothness of the function, favoring its usage in strictly real-time scenarios.
Recently, its use has been investigated in various applications, e.g., autonomous racing \cite{kabzan2019learning}, path following of off-road mobile robots \cite{ostafew2014learning}, and trajectory tracking with robotic manipulators \cite{carron2019data}.  In such applications, real-time feasibility is achieved with GPR applied in a grey-box context relying on a discrete-time learning-based formulation \cite{picotti2022continuous}.

In this manuscript, we present the application of a pure black-box LbNMPC controller to a scenario with complex dynamics, i.e., a go-kart platform, specifically designed to meet the real-time constraint. 
The proposed formulation eliminates the need for a priori knowledge of the dynamics by exploiting an entirely data-driven dynamical model that is defined in continuous time. Indeed, by conveniently adopting a continuous model of accelerations it is possible to directly use a space-based contouring problem formulation \cite{Lam2010}. This latter  computes the optimal trajectory and velocity profiles directly within the online optimization, requiring a sufficiently informative system model.
The continuous framework also allows the exploitation of a non-uniform integration grid, i.e. with  integration steps of different lengths within the prediction horizon, to reduce the number of model evaluations and limit the computational burden. 
To further limit the computational time, an ad-hoc GP model reduction strategy for GP-based LbNMPC has been implemented: specific approximations, based on the state prediction of the LbNMPC, are computed at every control step, leading to a sequence of low-dimensional local GP models. The article's objective is to describe the procedure, its advantages and disadvantages, mainly in relation to the trade-off between computational cost and performance.

A preliminary study to assess the feasibility of the proposed strategy and the tuning of the most relevant hyperparameters for real-time implementation has been conducted in a simulation scenario. Validation of the approach has been accomplished by driving a real go-kart vehicle (shown in Fig. \ref{fig:gokart}) on a $180m$ long indoor track. The strategies implemented to reduce computational burden allow for real-time execution of the LbNMPC controller on the embedded platform. Indeed, the proposed method allowed for accomplishing the driving task, properly performing the lap while keeping the go-kart within the track bounds, demonstrating that the black-box dynamic model can be a viable approach even in such a complex scenario. A comparison with an NMPC controller based on a validated nominal dynamic model has been also added, showing that the behaviors are comparable in terms of exhibited vehicle dynamics. Despite the satisfactory result, it is worth mentioning that the modest computational power of the system required the application of significant GP approximations, thus suggesting that performance improvements can be achieved by enhancing the computational capacity.

The rest of the paper is organized as follows. In Sec. \ref{sec:prob_stat} the control problem and the employed techniques are presented, while in Sec. \ref{sec:model} the LbNMPC internal model is described in detail. The mathematical description of the resulting optimal control problem is given in Sec. \ref{sec:lbmatmpc}, and Sec. \ref{sec:results} presents  the feasibility study in simulation an the obtained results on the real go-kart. Sec. \ref{sec:conclusion} draws the conclusions and presents future development.

\section{Problem Statement and Background}\label{sec:prob_stat}

In this manuscript, we present an LbNMPC controller based on a black-box model of the system, where the dynamics is obtained through GPR.
In the following, we describe the formulation of the unknown system within the LbNMPC framework and the GPR procedure adopted, as well as the reduction techniques used to obtain real-time performances.

\subsection{Model Definition}
Consider the following system 
\begin{equation}\label{eq:impl_dyn}
    \bm{\dot{x}}(t) = \bm{f}(\bm{x}(t),\bm{u}(t)), 
\end{equation}
where $\bm{x}\in\mathbb{R}^{n_x}$ is the system state and $\bm{u}\in\mathbb{R}^{n_u}$ is the system input. 
A physical description is usually used for deriving the model, hence the state can be divided as
\begin{equation}
    \bm{x} = [\bm{q}, \bm{\dot{q}}]^T,
\end{equation}
where $\bm{q} \in \mathbb{R}^{n_q}$ and $\bm{\dot{q}} \in \mathbb{R}^{n_q}$ are the position and velocities of the physical system, respectively.

While interacting with the system, the current system state $\bm{x}_k$, inputs $\bm{u}_k$ and \textit{real} accelerations $\bm{\ddot{q}}$ can be acquired for $T$ time steps. The data are assumed to be collected in discrete time instants, as usual in digital control, and the subscript $\cdot_{k}$ indicates the measurement at time instant $t_k$.
Group those data as
\begin{equation}\label{eq:data}
\begin{aligned}
    &\mathcal{X} = \{\bm{x}_1, \bm{x}_2, \dots , \bm{x}_{T} \} ,   \\
    &\mathcal{U} = \{\bm{u}_1, \bm{u}_2, \dots , \bm{u}_{T} \} ,   \\
    &\ddot{\mathcal{Q}} = \{\bm{\ddot{q}}_1, \bm{\ddot{q}}_2, \dots , \bm{\ddot{q}}_{T} \}.
\end{aligned}
\end{equation}

The quantities in \eqref{eq:data} can be used to estimate the \textit{acceleration functions} relying on GPR, as described in Sec. \ref{subsec:GPR}.
The obtained \textit{black-box} model of the system dynamics is expressed by the following equation
\begin{equation}\label{eq:overall_dyn}
\begin{aligned}
    \bm{\hat{\dot{x}}}(t) &= 
    \begin{bmatrix}
    \bm{\hat{\dot{q}}}(t) \\
    \bm{\hat{\ddot{q}}}(t)
    \end{bmatrix} &= 
    \begin{bmatrix}
    \bm{\dot{q}}(t) \\
    \bm{\psi}_{\ddot{q}}(t)
    \end{bmatrix} 
    \end{aligned},
\end{equation}
where $\bm{\psi_{\ddot{q}}}$ is the GP estimate of the accelerations and the hat $\hat{(\cdot)}$ symbol refers to predicted values. 

To conclude, define a numerical integrator operator 
\begin{equation}\label{eq:num_int}
\bm{\hat{x}}_{k+1} = \bm{\hat{\phi}}(\bm{x}_k,\bm{u}_k),    
\end{equation}
that integrates the dynamics $\bm{\hat{\dot{x}}}(t)$ and returns the solution at $t_{k+1}$ with $\bm{x}(0)=\bm{x}_k$, e.g. a 4th order Explicit Runge-Kutta.

\subsection{Gaussian Process Regression}\label{subsec:GPR}
Based on the data set in \eqref{eq:data}, an estimator of the dynamics can be derived relying on GPR. Each component of the acceleration vector $\bm{\ddot{q}}$ is modeled as a distinct and independent GP. The GPs input vector at time $t_k$, hereafter denoted with $\bm{x}^{gp}_k$, accounts for $\bm{x}_{k}$ and $\bm{u}_{k}$, namely, $\bm{x}^{gp}_k = \left[\bm{x}_{k}^T\,\bm{u}_{k}^T\right]^T$. The corresponding set of GPs input vector is referred to as $\mathcal{X}^{GP}$. The output of the \emph{i}-th GP is $y^i_k = \ddot{q}^i_k$, where $\ddot{q}^i_k$ is the measured \emph{i}-th component of $\bm{\ddot{q}}$. GPR considers the following probabilistic model
\begin{equation*}
    \bm{y}^i = \begin{bmatrix}
    y^i_1\\ \vdots \\ y^i_{T}
    \end{bmatrix} =
    \begin{bmatrix}
    \ddot{q}^i_1 \\ \vdots \\ \ddot{q}^i_T
    \end{bmatrix} =
    \begin{bmatrix}
    \bar{\psi}_{\ddot{q}}^i(\bm{x}^{gp}_1) \\ \vdots \\ \bar{\psi}_{\ddot{q}}^i(\bm{x}^{gp}_{T})
    \end{bmatrix}
    + 
    \begin{bmatrix}
    e^i_1 \\ \vdots \\ e^i_T
    \end{bmatrix} = \bar{\bm{\psi}}_{\ddot{q}}^i + \bm{e}^i,
\end{equation*}
where $\bm{e}^i$ is zero-mean independent Gaussian noise with standard deviation $\sigma_n$, while $\bar{\bm{\psi}}_{\ddot{q}}^i \sim N(0, K^i)$ is a zero-mean GP. A kernel function $k^i(\cdot,\cdot)$ defines the GP's covariance matrix $K^i$. 
Specifically, the covariance between $\bar{\psi}_{\ddot{q}}^i(\bm{x}^{gp}_k)$ and  $\bar{\psi}_{\ddot{q}}^i(\bm{x}^{gp}_j)$, i.e. the entry of $K^i$ at row $k$ and column $j$, is $E[\bar{\psi}_{\ddot{q}}^i(\bm{x}^{gp}_k) \bar{\psi}_{\ddot{q}}^i(\bm{x}^{gp}_j)] = k^i(\bm{x}^{gp}_k,\bm{x}^{gp}_j)$.
Let  $\bm{x}^{gp}_*$ be a general input location. Conveniently, the posterior distribution of $\bar{\psi}_{\ddot{q}}^i(\bm{x}^{gp}_*)$ is Gaussian, with mean 
and variance given by the following expressions \cite{rasmussen2003gaussian}:
\begin{align}
    &\psi_{\ddot{q}}^i(\bm{x}^{gp}_*) = \mathbf{k}_*^i \bm{\alpha}^i,\label{eq:GP_mean}\\
    &Var(\psi_{\ddot{q}}^i(\bm{x}^{gp}_*)) = k^i(\bm{x}^{gp}_*, \bm{x}^{gp}_*) - \mathbf{k}_*^i (K^i+\sigma_n^2 \mathbb{I})^{-1} (\mathbf{k}_*^i)^T, \label{eq:GP_var}
\end{align}
where 
\begin{align}\label{eq:k_alpha}
&\mathbf{k}^i_* = \begin{bmatrix}
k^i(\bm{x}^{gp}_*, \bm{x}^{gp}_1) & \dots k^i(\bm{x}^{gp}_*, \bm{x}^{gp}_T)
\end{bmatrix},\\
 & \bm{\alpha}^i = (K^i+\sigma_n^2 \mathbb{I})^{-1} \mathbf{y}^i,
\end{align}
with $\mathbb{I}\in \mathbb{R}^{T \times T}$ defined as the identity matrix. Then, since the posterior is Gaussian distributed, the maximum a posteriori estimate of $\bar{\psi}_{\ddot{q}}^i(\bm{x}^{gp}_*)$ is given by the posterior mean in \eqref{eq:GP_mean}, while the posterior variance \eqref{eq:GP_var} provides a confidence interval of the estimate. Consequently, the expression of $\bm{\psi}_{\ddot{q}}$ introduced in \eqref{eq:overall_dyn} is 
\begin{equation}\label{eq:psi_expr}
    \bm{\psi}_{\ddot{q}}(\bm{x}_{*},\bm{u}_{*}) = \begin{bmatrix}
    \psi_{\ddot{q}}^1(\bm{x}_{*},\bm{u}_{*})\\ \vdots \\ \psi_{\ddot{q}}^{n_q}(\bm{x}_{*},\bm{u}_{*})
    \end{bmatrix} = 
    \begin{bmatrix}
    \mathbf{k}^1_* \bm{\alpha}^1 \\ \vdots \\ \mathbf{k}^{n_q}_* \bm{\alpha}^{n_q}
    \end{bmatrix}.
\end{equation}
For future convenience, in the previous equation, we have made explicit the input of $\bm{\psi}_{\ddot{q}}$ instead of using $\bm{x}^{gp}_{k}$.

Regarding the kernel choice, several options are available, depending on a priori assumptions on $\bar{\psi}_{\ddot{q}}$. A convenient kernel for modeling continuous functions is the Squared Exponential (SE) kernel, 
defined as
\begin{align}
    k^i(\bm{x}^{gp}_k,\bm{x}^{gp}_j) =&
        e^{-(\bm{x}^{gp}_k-\bm{x}^{gp}_j)^T \Sigma^{i^{-2}} (\bm{x}^{gp}_k-\bm{x}^{gp}_j)} \nonumber\\
        &e^{-\norm{\bm{x}^{gp}_k - \bm{x}^{gp}_j }_{\mathcal{L}^i}^2}, \label{eq:SE-kernel}
\end{align}
where $\Sigma^{i}$ is a diagonal matrix, with the diagonal elements named lengthscales, and hereafter denoted $\mathcal{L}^i$. 
From the expression above we can see that the SE kernel defines the similarity between two samples based on the distance of their GP inputs. The set of GP's hyperparameters consists of the length scales and the noise standard deviation $\sigma_n$. The length scales determine the metric used to compute the distance between samples. Importantly, $\sigma_n$ determines the balance between adherence to the training data and complexity of $\bm{\psi}_{\ddot{q}}(\bm{x}_{k},\bm{u}_{k})$. High values of $\sigma_n$ promote the smoothness of $\bm{\psi}_{\ddot{q}}(\bm{x}_{k},\bm{u}_{k})$, but could limit accuracy. On the other hand, too small values of $\sigma_n$ could lead to overfitting and non-smooth $\bm{\psi}_{\dot{q}}(\bm{x}_{k},\bm{u}_{k})$, which might be an issue for the NMPC. The hyperparameters can be tuned relying on empirical methods, such as cross-validation, or by maximization of the training data Marginal Likelihood (ML), see \cite{rasmussen2003gaussian}. In this work, we selected the hyperparameters by optimizing the ML through a gradient-based optimization, exploiting the functionalities made available by \emph{pytorch} \cite{paszke2017pytorch}. 

\subsection{GP model approximation for real-time applications}\label{subsec:RTGP}
The main issue when dealing with learning dynamics LbNMPC is the computational burden induced by the data-driven model. As shown by \eqref{eq:GP_mean}, the time required for GP predictions grows linearly with the number of data points used. Besides that, the growth of the data points makes the model of the system dynamics more and more complex, leading to an optimization problem difficult to solve within the available control period. For these reasons, the reduction of the data points is fundamental in this framework. In this work, we implemented a strategy that, at each prediction time, aims at selecting a significant subset of the data points to limit the computational burden and dynamics complexity. We applied in cascade two reduction techniques. First, an \textit{offline} heuristic procedure, named Subset of Data (SoD), derives $\mathcal{X}^{GP}_{SoD}$, a subset of the original dataset $\mathcal{X}^{GP}$. Second, at execution time, a Nearest Neighbor (NN) \cite{datta2016hierarchical} approach selects $\mathcal{X}^{GP}_{NN}$, the subset of $\mathcal{X}^{GP}_{SoD}$ actually used for prediction. 
In this regard, the SE kernel introduced in \eqref{eq:SE-kernel} provides the NN algorithm with a principled metric to select the subset of nearest neighbor points. Furthermore, it approximates well functions close to training inputs, which is the goal of our local models. 
In the following, we describe the two reduction techniques implemented for each GP acceleration model.

\subsubsection{Subset of Data}
the heuristic procedure implemented to reduce the original dataset $\mathcal{X}^{GP}$ consists of an iterative algorithm that runs \emph{offline} and exploits information provided by the posterior variance \eqref{eq:GP_var}. First, $\mathcal{X}^{GP}_{SoD}$ is initialized with the first data point of $\mathcal{X}^{GP}$. Then, the algorithm iterates over the remaining points in $\mathcal{X}^{GP}$. For each point, the algorithm computes \eqref{eq:GP_var} using as training data the current points in $\mathcal{X}^{GP}_{SoD}$, and includes the new point in $\mathcal{X}^{GP}_{SoD}$ if the variance is higher than a certain threshold provided by the user. The basic idea behind this approach is that if the model is confident in the current input location we can neglect this point, while if the variance is high we add the point to $\mathcal{X}^{GP}_{SoD}$ to improve prediction accuracy. 

\subsubsection{Nearest Neighbor}
this approach is based on the key idea that the closest points to the target are the most informative for the prediction of the GP. In particular, the closest points are defined by the distance considering the trained lengthscales $\mathcal{L}$ using a weighted norm, i.e. $d_{k,j} = \norm{\bm{x}^{gp}_k - \bm{x}^{gp}_j }_{\mathcal{L}}^2$. The search for the \textit{nearest points} is done on the previously obtained inducing set $\mathcal{X}^{GP}_{SoD}$, obtaining a further reduced number of inducing points $T^{NN}<T^{SoD}<T$. Once the points have been selected, the quantities to obtain the posterior mean of the GP have to be recomputed, i.e. $\psi_{\ddot{q}}^{NN}(\bm{x}^{gp}_*) = \mathbf{k}_*^{NN} \bm{\alpha}^{NN}$, where $\mathbf{k}_*^{NN}$ and $\bm{\alpha}^{NN}$ are defined as in \eqref{eq:k_alpha}, where $T = T^{NN}$ and $\mathbf{y}=\mathbf{y}^{NN}$ is the subset of measures related to the chosen points. This procedure is accomplished online before the LbNMPC call, selecting, for each step in the prediction horizon, the closest points with respect to the trajectory predicted at the previous iteration, and updating the GP-related parameters within the Optimal Control Problem (see Sec. \ref{sec:lbmatmpc}).

\section{Continuous LbNMPC Model}\label{sec:model}

The prediction model of the LbNMPC is presented hereafter. 
The dynamics are completely data-driven, i.e. both the direct lateral acceleration $\dot{v}_y$ and yaw acceleration $\ddot{\theta}$ are entirely characterized by two different Gaussian Processes. The longitudinal command of the go-kart is based directly on the requested longitudinal acceleration, i.e. the desired acceleration is transmitted to the control system of the electric motors on the wheels that is responsible for actuation, so the characterization of the longitudinal acceleration is not needed.
Both simulation and experimental scenarios are considered, in order to evaluate the strategy also in absence of noise and actuator dynamics. 
On the other hand, the spatial domain reformulation has been mathematically described, characterizing the velocity rotations, in order to define a time-minimization strategy. Finally, a previously derived nominal dynamics model is presented for comparison.

\subsection{Black-box Dynamics Model}
The black-box dynamics model has been obtained as detailed in Sec. \ref{subsec:GPR} and applying the approximations described in Sec. \ref{subsec:RTGP}. In particular, the go-kart state components $\bm{x}^{gp} = [v_x, v_y, \dot{\theta}, \gamma, \beta, \tau_v]$, i.e., vehicle velocities and commands (see Sec. \ref{sec:complete_model}), have been used as regressors for the GP training, and a minimum $\sigma_n = 0.15$ has been imposed to enhance regularization properties, both for simulation and experimental scenarios. The GP has then been trained in batch, using a batch size of $100$ data points, for $400$ epochs and a learning rate of $0.001$. Furthermore, SoD reduction has been applied with threshold $1.0\sigma_n^2$. To check the prediction capabilities of the method that will be employed online, the NN strategy with $T^{NN}_{\dot{v}_y} = 30$ and $T^{NN}_{\ddot{\theta}} = 50$ for $\dot{v}_y$ and $\ddot{\theta}$, respectively, has been applied on the obtained GP. Such dimensions, in fact, allow balancing between the computational burden and prediction capabilities of the controller. Moreover, the nominal model (described in Sec. \ref{subsec:nominal_model}) has been tested on the same data.

\begin{table}[b]
    \centering
    \caption{RMSE of lateral and yaw accelerations for nominal and black-box models (both using SoD and NN reductions) in simulation.}
    \label{tab:sim_comp_err}
    \begin{tabular}{lcc}
        model & $\dot{v}_y$ $[m/s^2]$ & $\ddot{\theta}$ $[rad/s^2]$  \\
        \hline\\
        nominal & 0.62 & 1.10 \\
        black-box SoD & 0.44 & 0.73 \\
        black-box NN & 0.61 &  0.74
    \end{tabular}
\end{table}

\subsubsection{Simulation}
\label{subsubsec:simulation_GP}
due to the simulation environment structure, a nominal controller-driven run, i.e. using the nominal model within the NMPC controller (see Sec. \ref{sec:lbmatmpc}), was recorded and used for black-box model training. 
As the simulation serves as a preliminary stage for the experimental setup and manual driving was not possible due to hardware limitations, we used a closed-loop control action to mimic human driving.
This step allows for verifying the capability of the GP to fit the acceleration data in a fully controlled environment and comparing the accuracy with respect to the nominal model. The sparse GP after SoD reduction contains 53 inducing points for both $\dot{v}_y$ and $\ddot{\theta}$, hence the local approximation with $T^{NN}$ points precision is almost similar.
The Root Mean Squared Error (RMSE) of both lateral and yaw accelerations for nominal and black-box dynamics resulted to be very similar (reported in Tab. \ref{tab:sim_comp_err}), hence supporting the possibility to use the developed strategy for the real platform.

\subsubsection{Experimental}\label{subsubsec:model_exp}

\begin{figure}[t]
    \centering
    \includegraphics[width=0.99\linewidth]{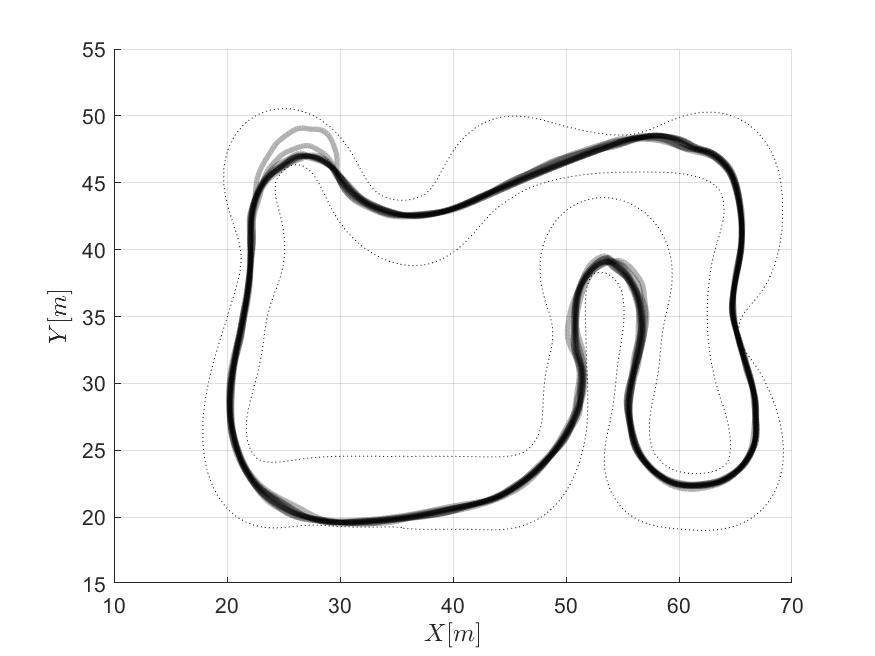}
    \caption{Trajectories of the data acquired for training.}
    \label{fig:data_XY}
\end{figure}
\begin{figure}[t]
    \centering
    \includegraphics[width=0.9\linewidth]{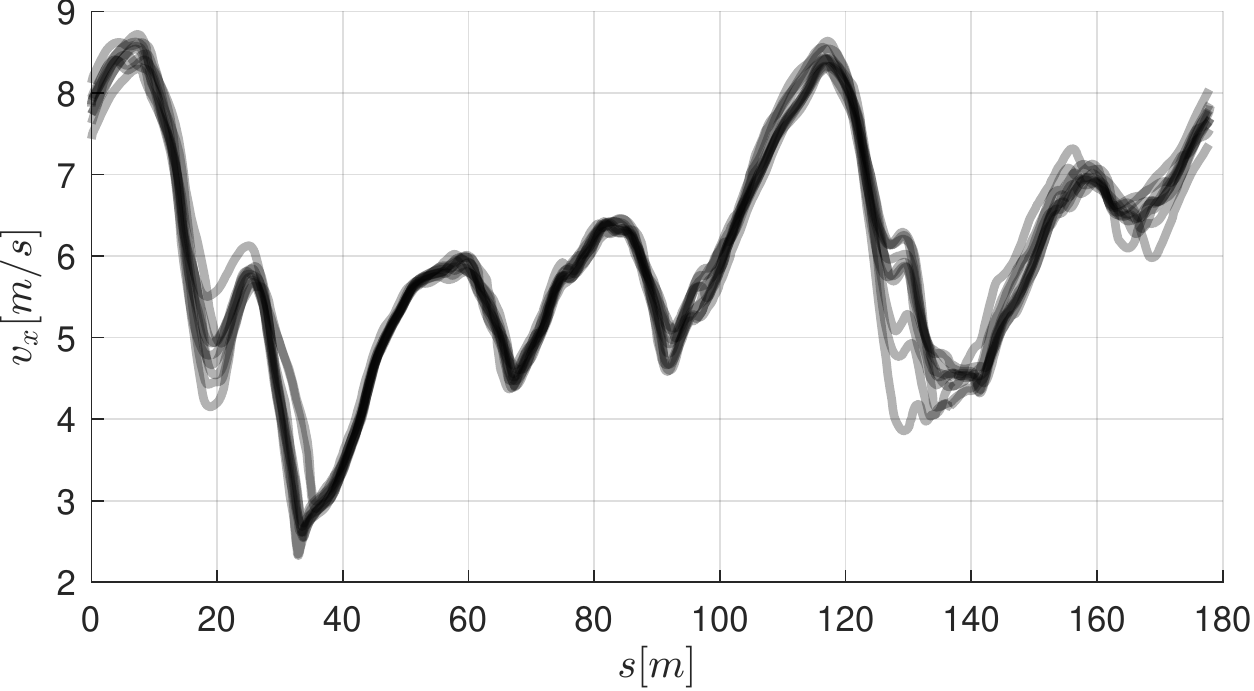}
    \caption{Longitudinal velocity of the data acquired for training.}
    \label{fig:data_vx}
\end{figure}
\begin{figure}[t]
    \centering
    \includegraphics[width=0.9\linewidth]{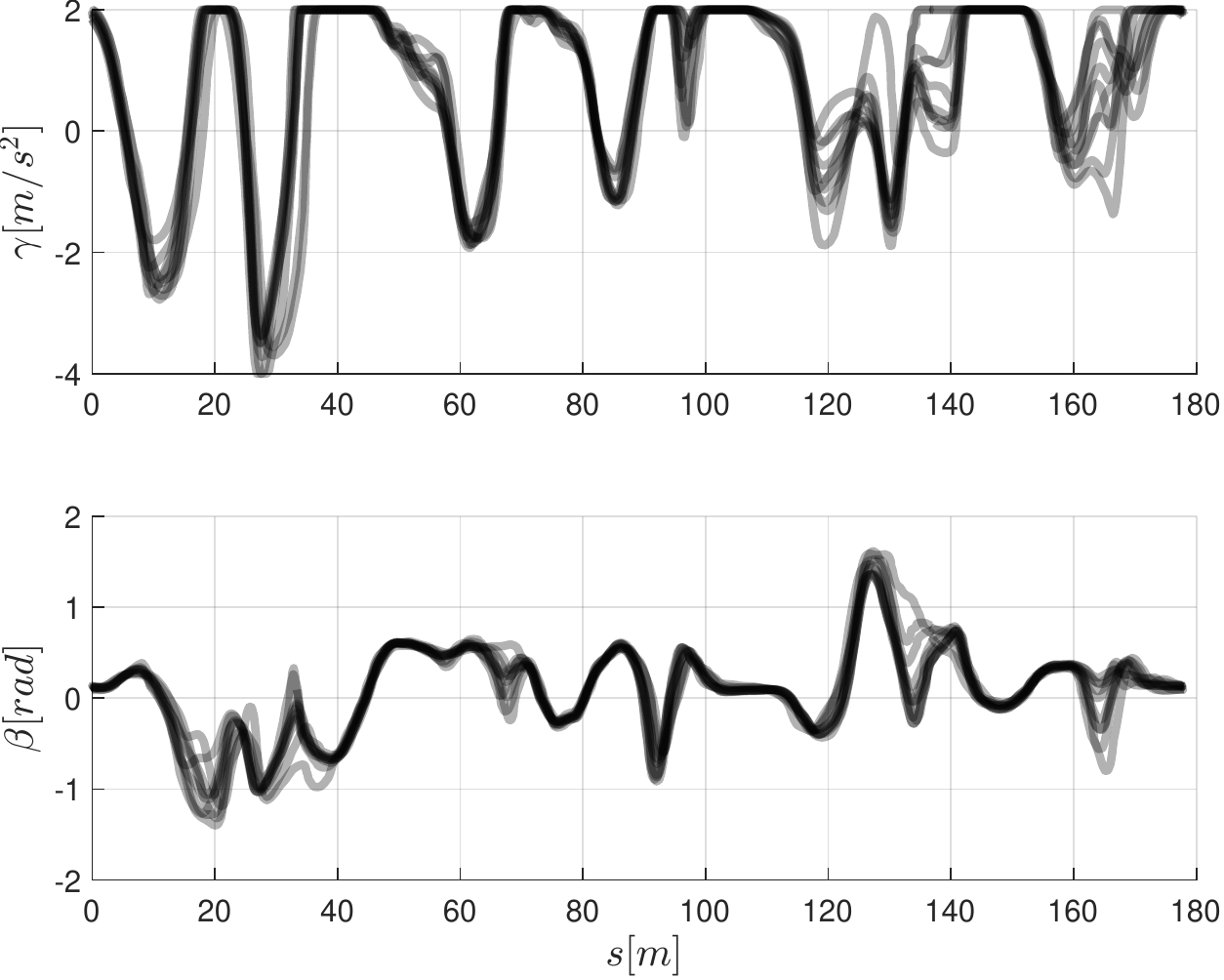}
    \caption{Commands given to the go-kart of the data acquired for training.}
    \label{fig:data_cmd}
\end{figure}
\begin{figure}[t]
    \centering
    \includegraphics[width=0.9\linewidth]{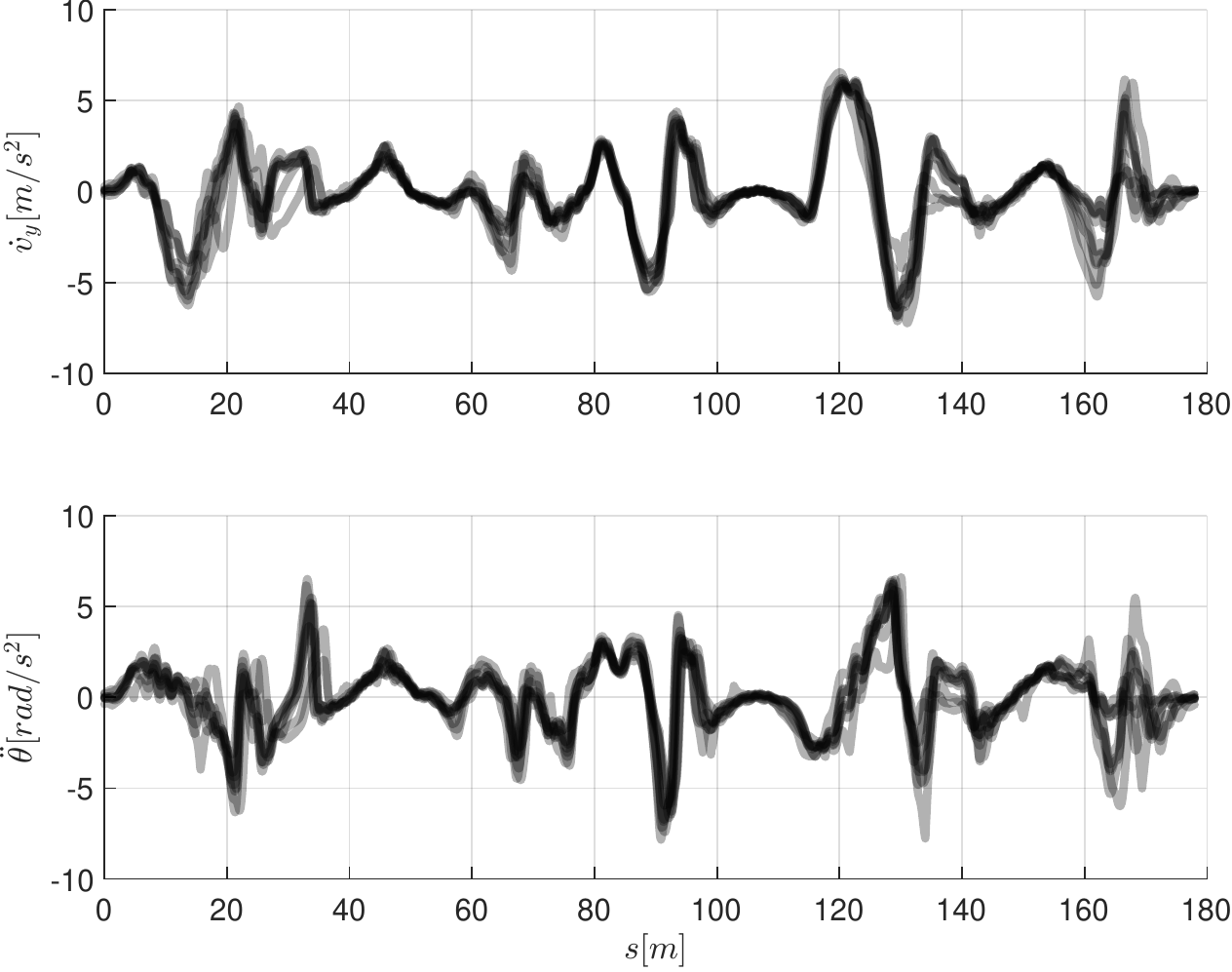}
    \caption{Direct lateral acceleration and yaw acceleration of the data acquired for training.}
    \label{fig:data_accs}
\end{figure}

\begin{figure}[t]
    \centering
    \includegraphics[width=0.9\linewidth]{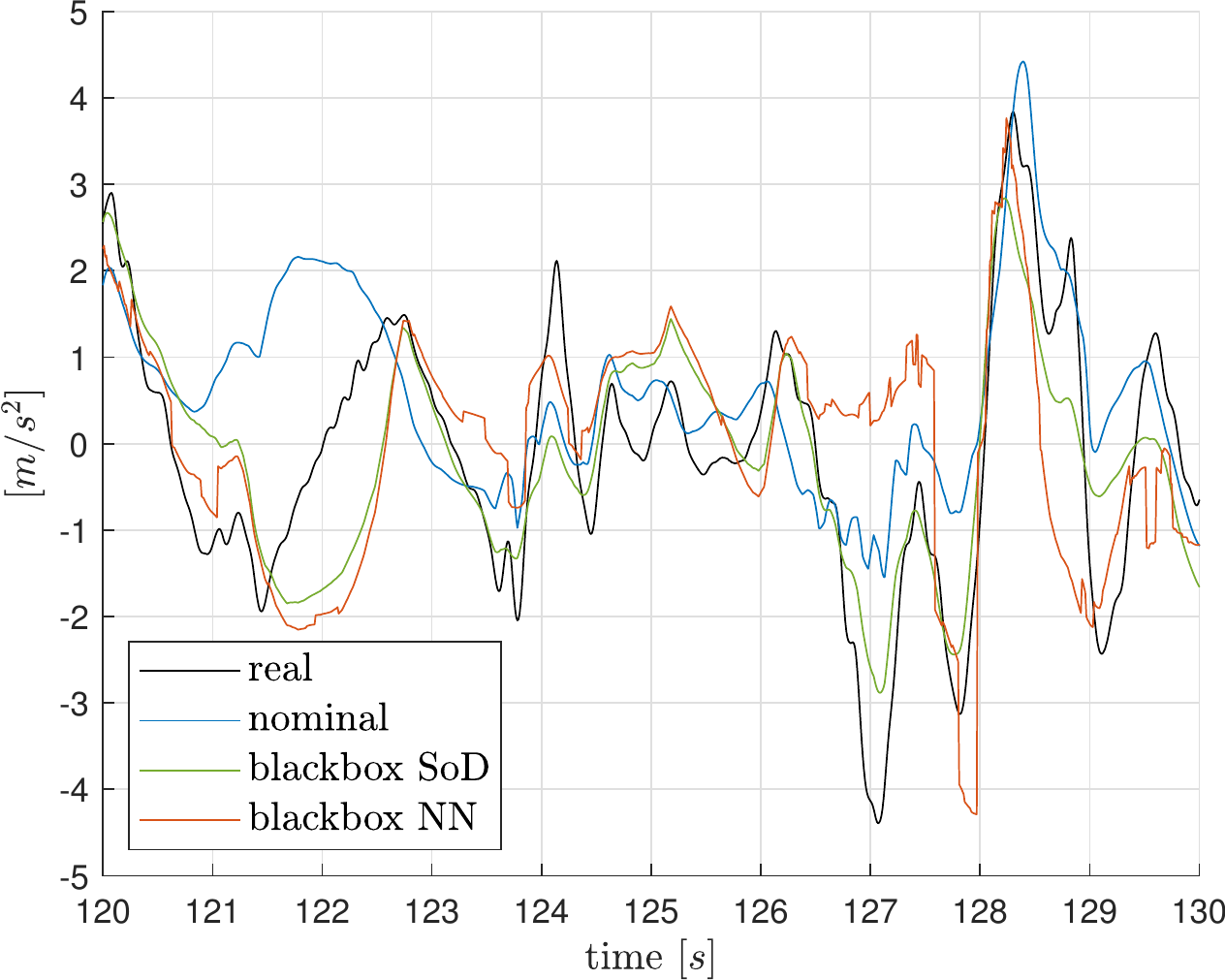}
    \caption{Comparison of nominal and black-box models (both with SoD and NN reductions) with respect to the real \textit{direct} lateral acceleration $\dot{v}_y$ on the real go-kart.}
    \label{fig:cmp_vydot}
\end{figure}
\begin{figure}[t]
    \centering
    \includegraphics[width=0.9\linewidth]{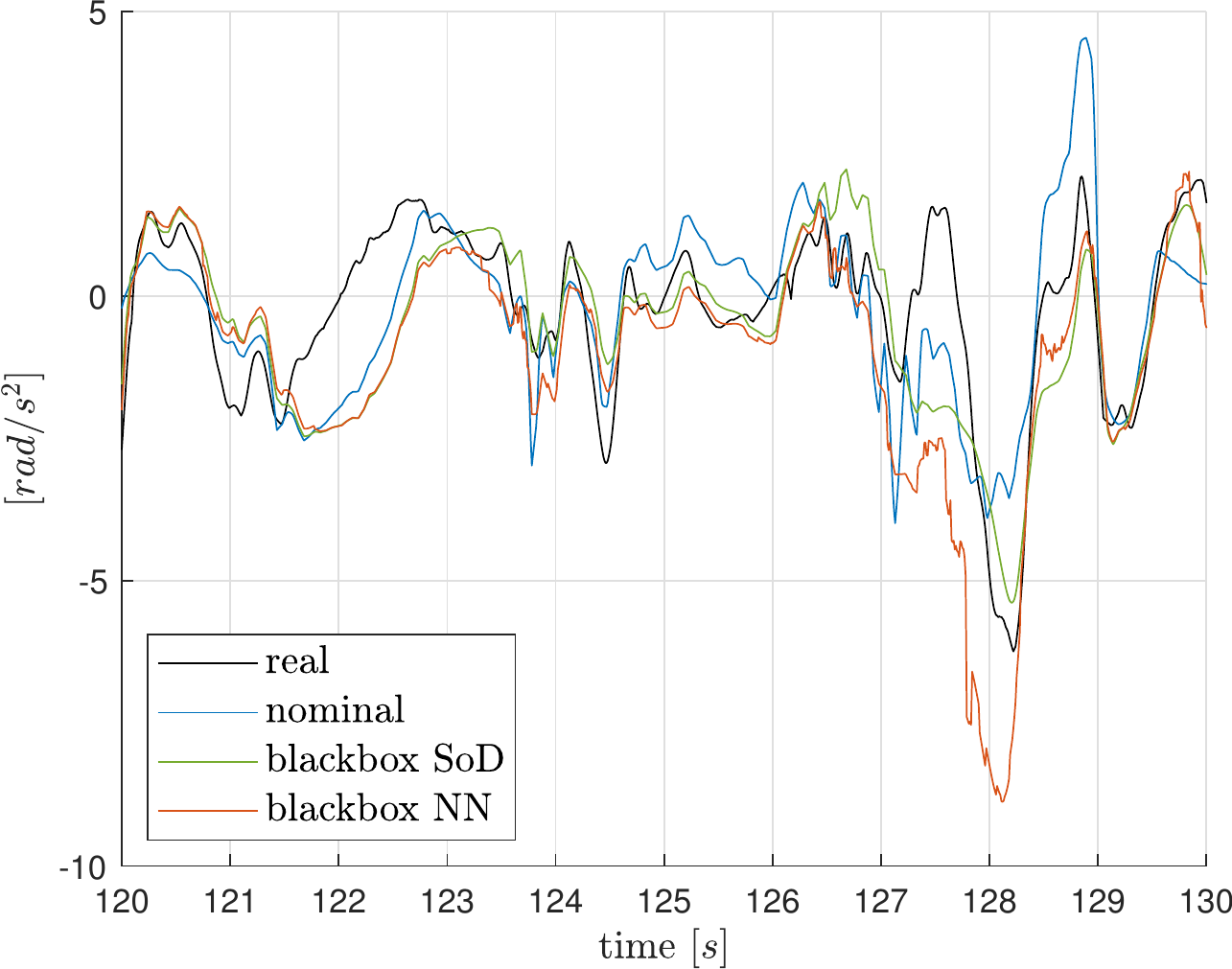}
    \caption{Comparison of nominal and black-box models (both with SoD and NN reductions) with respect to the real yaw acceleration $\ddot{\theta}$.}
    \label{fig:cmp_thetaddot}
\end{figure}
to excite the highly nonlinear and complex dynamics of the system, 10 human-driven laps close to the vehicle's maximum acceleration were performed. A total number of around $5000$ data points was recorded, comprising IMU, localization, and velocity data, along with steering position and longitudinal acceleration commands. 
A depiction of the acquired data in terms of path, velocity profiles and used commands is reported in Fig. \ref{fig:data_XY}-\ref{fig:data_cmd}. The travelled trajectories are quite similar, as expected in a track driving task, but the different laps allow to give sufficient variety in the training data.
Additionally, it is worth mentioning that GPs do not include the curvilinear abscissa as a predictor variable. Instead, they are capable of estimating accelerations by exploiting the system's velocities and inputs, which separates the vehicle's dynamic behavior from its position on the track. As a result, the various turns on the track present different operating conditions for the system, making the 10 laps a sufficiently information-rich training dataset.
The resulting data, acquired at different frequencies depending on the sensor, were interpolated and the IMU and velocity data were filtered through a Forward Backward Kalman Filter \cite{BFKF}, resulting in aligned and smooth acceleration data for the GP training,
shown in Fig. \ref{fig:data_accs}.
SoD reduction led to sparse GPs of $98$ and $114$ points for $\dot{v}_y$ and $\ddot{\theta}$, respectively.
The resulting accelerations estimates, i.e. the mean of the GP for lateral $\psi_{\dot{v}_y}$ and yaw $\psi_{\ddot{\theta}}$ accelerations, are shown in Fig. \ref{fig:cmp_vydot} and \ref{fig:cmp_thetaddot}, together with the real accelerations on the go-kart. Both the ability of the model to fit the data and the approximation given by the reduction are clearly visible. In particular, the NN strategy on a 6-dimensional space is not always effective and may generate some spikes. 
At the same time, note that also the nominal model is not able to always describe correctly the acceleration values, but mostly the tendency.
However, the Root Mean Squared Error (RMSE) of lateral and yaw accelerations, reported in Tab. \ref{tab:comp_err}, is still comparable with the nominal dynamics ones. 

\begin{table}[tb]
    \centering
    \caption{RMSE of lateral and yaw accelerations for nominal and black-box models (both using SoD and NN reductions) on the real go-kart.}
    \label{tab:comp_err}
    \begin{tabular}{lcc}
        model & $\dot{v}_y$ $[m/s^2]$ & $\ddot{\theta}$ $[rad/s^2]$  \\
        \hline\\
        nominal & 1.68 & 1.28 \\
        black-box SoD & 1.03 & 1.22 \\
        black-box NN & 1.95 & 1.81 
    \end{tabular}
\end{table}

\subsection{Spatial Reformulation}

\begin{figure}[tb]
	\centering
	\includegraphics[width=.95\linewidth]{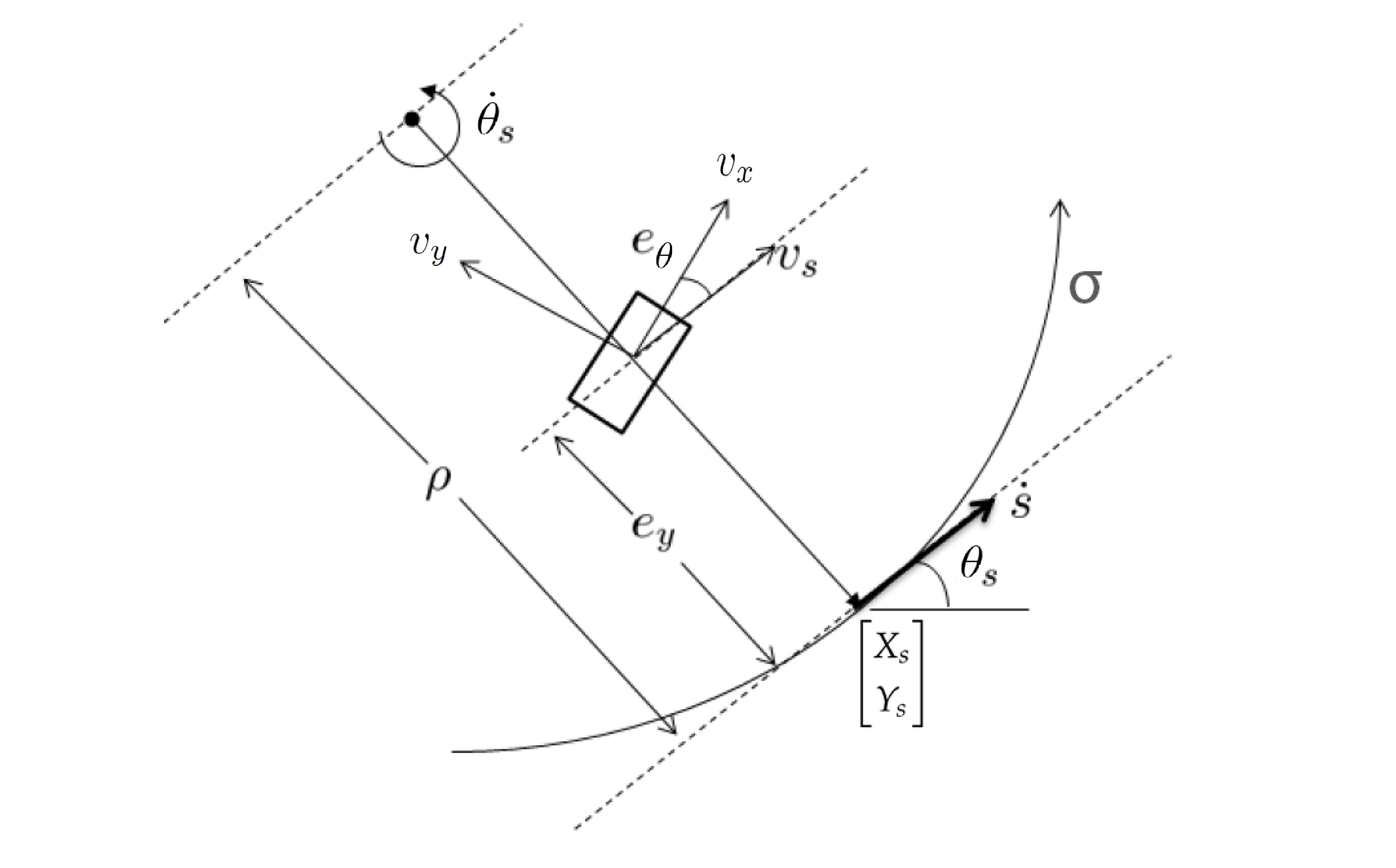}
	\caption{The spatial coordinates system. It allows defining the lateral and heading errors and formulating the dynamics wrt the curvilinear abscissa $s$.}\label{fig:s-coord}
\end{figure}

The complete model is comprehensive of the yaw $e_\theta$ and lateral $e_y$ errors with respect to the track centerline and the integration has been applied in spatial coordinates with respect to the arc length $s$ along the track, as shown in Fig. \ref{fig:s-coord} and previously presented in \cite{brus2019driver,picotti2022nonlinear}. This strategy allows for using time as a minimization variable, setting the spatial constraints, and eliminating the dependency on the velocity in the trajectory reference for the controller. The kinematic relation between velocities and the angular and lateral errors can be expressed as
\begin{equation}
\begin{aligned}
	\dot{e}_{\theta} &= \dot{\theta} - \zeta \ \dot{s}, \\
	\dot{e}_y &= v_x \ \mathrm{sin}(e_{\theta}) + v_y \ \mathrm{cos}(e_{\theta}),
\end{aligned}
\end{equation}
where $v_x, v_y, \dot{\theta}$ are the longitudinal, lateral, and yaw velocities, respectively, $\zeta=\frac{1}{\rho}$ is the trajectory curvature, and $\dot{s} = \frac{1}{1 - \zeta \ e_y} \left(\dot{x} \ \mathrm{cos}(e_{\theta}) - \dot{y} \ \mathrm{sin}(e_{\theta}) \right)$. 

\subsection{Complete Model}\label{sec:complete_model}
The problem has been formulated in \textit{velocity form}, i.e., the inputs are the derivatives of the actual go-kart commands. Hence, the state has been defined as
\begin{equation}
    \bm{x} =
    [
    v_x, \ v_y, \ \dot{\theta}, \ e_\theta, \ e_y, \ \gamma, \ \beta, \ \tau_v, \ t
    ]^T,
\end{equation}
where $\gamma$, $\beta$, and $\tau_v$ are the actual commands to the go-kart, i.e. the desired longitudinal acceleration, the steering angle, and the torque vectoring component, respectively, and $t$ is the time, while the input vector is
\begin{equation}
    \bm{u} =
    [
    \dot{\gamma}, \ \dot{\beta}, \ \dot{\tau}_v, \ \eta
    ]^T,
\end{equation}
where $\dot{\gamma}, \ \dot{\beta}, \ \dot{\tau}_v$ are the derivatives of the actual commands, and $\eta$ is a slack variable needed for implementing a soft constraint \cite{Zeilinger2010}.
The resulting dynamics are expressed by
\begin{equation}
    \dot{\bm{x}} =
    [
    \gamma, \ \psi_{\ddot{y}}, \ \psi_{\ddot{\theta}}, \ \dot{e}_\theta, \ \dot{e}_y, \ \dot{\gamma}, \ \dot{\beta}, \ \dot{\tau}_v, \ 1 
    ]^T.
\end{equation}
Finally, the state vector $x$ is differentiated w.r.t $s$ using the \textit{chain rule} as
\begin{equation}
    \bm{x}' = \frac{d\bm{x}}{ds} = \frac{d\bm{x}}{dt} \frac{dt}{ds} = \frac{d\bm{x}}{dt} \frac{1}{\dot{s}} = \frac{\dot{\bm{x}}}{\dot{s}}, \, \, \, \forall \dot{s} \neq 0,
\end{equation}
thus defining the integration with respect to the curvilinear abscissa $s$.

\subsection{Nominal Model}\label{subsec:nominal_model}

A complete dynamics model was already available and has been used for comparison. It is based on a three-wheel vehicle and includes different important dynamics, such as 
\begin{itemize}
    \item an identified Pacejka's magic formula for lateral tire forces,
    \item an approximation of the adherence ellipsoid for combined tire behavior,
    \item longitudinal load transfer dynamics.
\end{itemize}
In particular, the Pacejka's formulas for front and rear wheels are the core of the model, as they require specific tests to correctly identify the shape of the curves.
Further details on the formulation can be found in \cite{mion2021phd}.

\section{LbNMPC Formulation}\label{sec:lbmatmpc}

The NMPC problem has been implemented in \software{LbMATMPC}, an open-source software developed in \software{MATLAB} for the development of NMPC strategies \cite{picotti2022lbmatmpc}, and easily exploitable for C implementation.

At every time instant $k$, the OCP is converted in a NLP through direct multiple shooting methods over the prediction horizon, which is composed by $N$ \emph{shooting intervals} $[t_{0|k},t_{1|k},\ldots,t_{N|k}]$, as follows: 
\begin{subequations}\label{NLP}
\begin{align}
\min_{\bm{x},\bm{u}} \,&\sum_{{j}=0}^{N-1} \frac{1}{2}\norm{\bm{h}_{j}(\bm{x}_{j|k},\bm{u}_{j|k})}_W^2+\frac{1}{2}\norm{\bm{h}_N(\bm{x}_{N|k})}_{W_N}^2 \label{cost function}\\
s.t.\, &0=\bm{x}_{0|k}-\bar{\bm{x}}_{0|k},\label{initial value embedding}\\
&0=\bm{x}_{j+1|k}-\bm{\hat{\phi}}(\bm{x}_{j|k},\bm{u}_{j|k}),\label{conti const}\,j=0,1,\ldots,N-1,\\
&\underline{\bm{r}}_{j|k}\leq \bm{r}(\bm{x}_{j|k},\bm{u}_{j|k})\leq \overline{\bm{r}}_{j|k}, \,j=0,1,\ldots,N-1,\label{path constraint}\\
&\underline{\bm{r}}_{N|k}\leq \bm{r}_N(\bm{x}_{N|k})\leq \overline{\bm{r}}_{N|k},\label{final path constraint}
\end{align}
\end{subequations}
where $\bar{\bm{x}}_{0|k}$ is the state at the current time instant $k$. At the discrete time point $t_{j|k}$ for $j=0,\ldots,N$ the system states $\bm{x}_{j|k}\in\mathbb{R}^{n_x}$ are defined while the control inputs $\bm{u}_{j|k}\in \mathbb{R}^{n_u}$ for $j=0,\ldots,N-1$ are piece-wise constant. Eq. \eqref{cost function} refers to the objective function in which the inner objectives $\bm{h}_j$ and $\bm{h}_N$ are expressed in Eq. \eqref{eq:cf}. Eq. \eqref{initial value embedding} refers to the initial value embedding and the constraint function \eqref{path constraint}-\eqref{final path constraint} is defined as $\bm{r}(\bm{x}_{j|k},\bm{u}_{j|k}): \mathbb{R}^{n_x}\times\mathbb{R}^{n_u} \rightarrow \mathbb{R}^{n_r}$ and $\bm{r}_N(\bm{x}_{N|k}): \mathbb{R}^{n_x}\rightarrow \mathbb{R}^{n_{r_N}}$ with upper and lower bound $\overline{\bm{r}}_{j|k}, \underline{\bm{r}}_{j|k}$. The learning-based dynamics, Eq. \eqref{eq:num_int}, is enforced by the \emph{continuity constraint}, Eq. \eqref{conti const} \cite{bock1984multiple}.

SQP method is used to solve problem \eqref{NLP}, i.e. through an iterative procedure that reformulates the NLP at a given iterate in a Quadratic Programming (QP) problem. Specifically, the objective functions \eqref{cost function} are replaced by their local quadratic approximation and the constraint functions \eqref{initial value embedding}-\eqref{final path constraint} by their local affine approximations. 
Thus, at SQP iterate $l$, a QP problem is formulated as follows (the subscript $ _{\cdot|k}$ is omitted for clarity): 

\begin{align}\label{QP}
\begin{split}
\min_{\Delta \mathbf{x},\Delta \mathbf{u}} \quad & \sum_{j=0}^{N-1}( \frac{1}{2}
\begin{bmatrix}
\Delta \bm{x}_j\\
\Delta \bm{u}_j
\end{bmatrix}^\top H^l_j 
\begin{bmatrix}
\Delta \bm{x}_j\\
\Delta \bm{u}_j
\end{bmatrix} + g_j^{i^\top}
\begin{bmatrix}
\Delta \bm{x}_j\\
\Delta \bm{u}_j
\end{bmatrix} ) \\
& + \frac{1}{2}\Delta \bm{x}_N^\top H^l_N \Delta \bm{x}_N + g_N^{l^\top}\Delta \bm{x}_N \\
s.t. \quad & \Delta \bm{x}_0=\hat{\bm{x}}_0-\bm{x}_0,\\
& \Delta \bm{x}_{j+1}= A_{j}^l \Delta \bm{x}_{j} + B_{j} \Delta \bm{u}_{j} +\bm{a}_{j}^l,\\
& \underline{\bm{c}}_j^l\leq C_j^l \Delta \bm{x}_j + D_j^l \Delta \bm{u}_j\leq \overline{\bm{c}}_j^l, \\
&\underline{\bm{c}}_N^l\leq C_N^l \Delta \bm{x}_N \leq \overline{\bm{c}}_N^l, 
\end{split}
\end{align}
where $\Delta \mathbf{x} =\mathcal{x}-\mathcal{x}^l, \Delta \mathbf{u}=\mathcal{u}-\mathcal{u}^l$, using the compact notation $
\mathcal{x}= \left [\bm{x}_0^\top, \bm{x}_1^\top,\dots, \bm{x}_N^\top\right ]^\top, \,
\mathcal{u}= \left [\bm{u}_0^\top, \bm{u}_1^\top,\dots, \bm{u}_{N-1}^\top\right ]^\top
$
for the discrete state and control variables, and $\mathcal{x}^l$ and $\mathcal{u}^l$ are the previous guess for state and control trajectories. The linearization matrices are given by
\begin{equation}\label{QP data}
\begin{aligned}
&A_{j}^l=\frac{\partial \bm{\hat{\phi}}}{\partial \bm{x}_j}, \quad B_{j}^l=\frac{\partial \bm{\hat{\phi}}}{\partial \bm{u}_j},\\
&\bm{a}_{j}^l = \bm{\hat{\phi}}(\bm{x}_j^l,\bm{u}_j^l) -\bm{x}_{j+1}^l,\\
&C_j^l=\frac{\partial \bm{r}_j}{\partial \bm{x}_j}, \quad D_j^l=\frac{\partial \bm{r}_j}{\partial \bm{u}_j},\quad C_N^l=\frac{\partial \bm{r}_N}{\partial \bm{x}_N},\\
&\overline{\bm{c}}_j^l=\overline{\bm{r}}_j-\bm{r}_j(\bm{x}_j^l,\bm{u}_j^l),\quad  \underline{\bm{c}}_j^l=\underline{\bm{r}}_j-\bm{r}_j(\bm{x}_j^l,\bm{u}_j^l),\\
&\overline{\bm{c}}_N^l=\overline{\bm{r}}_N-\bm{r}_N(\bm{x}_N^l),\quad  \underline{\bm{c}}_N^l=\underline{\bm{r}}_N-\bm{r}_N(\bm{x}_N^l).
\end{aligned}
\end{equation}
and the Hessian matrices $H_j^l, H_N^l$ are obtained by the constrained Gauss-Newton
method \cite{diehl2002real}.

The solution of \eqref{QP} is then used to update the solution of \eqref{NLP} by
\begin{equation}
\begin{aligned}
    \mathcal{x}^{l+1} &= \mathcal{x}^{l} + \beta^l \Delta\mathbf{x}^{l}, \\ 
    \mathcal{u}^{l+1} &= \mathcal{u}^{l} + \beta^l \Delta\mathbf{u}^{l},
\end{aligned}
\end{equation}
where $\beta^l$ is the step length determined by globalization strategies. 
Additional details on the algorithm can be found in \cite{chen2019matmpc, picotti2022lbmatmpc}.

\section{Results}\label{sec:results}
In this section, the results obtained are presented: a preliminary simulative trial allows the viability of the procedure to be analyzed in a fully controlled environment, while the experimental scenario illustrates the actual validation of the strategy. The controller must determine both the trajectory and the velocity profile in the prediction horizon, knowing the track bounds, while the time-minimization is enforced in the cost function. This formulation allows adapting to unexpected behaviors of the vehicle since the desired path and speed are recomputed at every time step.
The prediction capabilities are also investigated, considering the one-step ahead velocity prediction error, i.e. $\hat{e}_{\dot{q}} = \hat{\dot{q}} - \dot{q}$, where $\hat{\dot{q}}$ is the one-step-ahead velocity prediction and $\dot{q}$ is the real velocity of the go-kart. To fulfill hard real-time requirements, local approximations (see Sec. \ref{subsec:RTGP}) and non-uniform grid integration (see Sec. \ref{subsec:ctrl_setup}) have been adopted.

Through the whole section, the proposed LbNMPC based on black-box modeling and an NMPC based on nominal dynamics are compared, referring to the two controllers as \textit{black-box} and \textit{nominal} strategy, respectively.

\subsection{Controller Setup}\label{subsec:ctrl_setup}

The cost function for the LbNMPC is defined as
\begin{equation}\label{eq:cf}
\begin{aligned}
    \bm{h}_j(\bm{x}_j,\bm{u}_j) =  [&\dot{\gamma}, \dot{\tau}_v, \dot{\beta}, \eta]^\top, \\
    \bm{h}_N(\bm{x}_N) = [& t, e_{\psi} - e_{\psi}^{\mathrm{ref}}, e_y - e_y^{\mathrm{ref}}]^\top.
\end{aligned}
\end{equation}
The input terms allow a smooth control action, while the objective variable time $t$ supports the computation of a time-minimizing path and its weight can be used to tune the importance of the lap time performance. The slack variable $\eta$ is used to define the soft constraint on the track bounds violations.
The terminal cost terms related to errors $e_{\theta} \text{ and } e_y$ are adopted to enforce reasonable dynamics of the vehicle at the end of the prediction horizon. The trajectory to be used for this task has been precomputed by minimizing the path curvature.
To maintain the same configuration in all the tests and to allow fair comparisons of the results in the different cases, the controller tuning has been accomplished through an empirical analysis, resulting in the following weight matrices
\begin{equation}
\begin{aligned}
    W &= diag([2 \cdot 10^{-3}, 5 \cdot 10^{-2}, 10^{-2}, 5 \cdot 10^{1}]),\\
    W_N &= diag([10^{-1}, 10^{3}, 10^{2}]).
\end{aligned}
\end{equation}
The constraints are defined as 
\begin{equation}
\begin{aligned}\label{eq:constraints}
    \bm{r}_j = [&v_x, e_\theta, \gamma, \beta, \tau_v, \dot{\gamma}, \dot{\tau}_v, \dot{\beta}, \eta, e_y + \eta]^\top, \\
    \bm{r}_N = [&v_x, e_\theta, \gamma, \beta, \tau_v, e_y + \eta]^\top,
\end{aligned}
\end{equation} 
where the constraints on $v_x$ and $e_\theta$ are used to exclude singularities in the model kinematics and numerical issues, the ones on the states $\gamma, \ \beta \text{ and } \tau_v$ are the integration of the computed inputs and represent intrinsic bounds of the actual vehicle commands, while those on $\dot{\gamma}, \ \dot{\beta} \text{ and }\dot{\tau}_v$ are added in order to improve the smoothness of the computed inputs and can be used to tune the aggressivity of the NMPC driving commands. Finally, the constraint on $\eta$ allows setting a maximum track bound exceed and the one on $e_y$ defines the width of the track.
The bounds are hence defined as
\begin{equation}\resizebox{0.95\hsize}{!}{$
\begin{aligned}
\underline{\bm{r}} = [2.5, -\pi/2, -4.2, -\pi/2, -1.7, -10^3, -10^2, -10^1, -5, e_y^{lb}]^T, \\
\overline{\bm{r}} = [15, +\pi/2, +2, +\pi/2, +1.7, +10^3, +10^2, +10^1, +5, e_y^{ub}]^T,
\end{aligned}$}
\end{equation} 
where $e_y^{lb}$ and $e_y^{ub}$ are the track bounds updated online based on the current position, slightly reduced by $0.5m$ to effectively implement the soft constraint.

HPIPM \cite{frison2020hpipm} has been used within LbMATMPC as QP sparse solver and Real-Time Iteration (RTI) scheme \cite{gros2020from} has been adopted. By using the RTI scheme, local convergence of the algorithm is ensured by RTI guarantees on contractivity and boundness of the loss of optimality compared to optimal feedback control \cite{diehl2005real,gros2020from}. The integration step has been set as $T_s = 0.3m$ for $N=80$ steps, allowing a prediction horizon of $24m$, and a reduction of the shooting points has been obtained through a non-uniform integration grid of $r = 33$ points distributed as $G = $
$\{1, 2, 3, 4, 5, 6, 8, 10, 12, 14, 16, 18, 20, 23, 26, 29, 32, 35, 38,$
$41, 44, 47, 50, 53, 56, 59, 62, 65, 68, 71, 74, 77, 80\}$.
The grid allows accurate control actions to be computed at the beginning of the horizon while reducing the computational burden and enabling less precision for subsequent points in the future. The technique, with the number of data points chosen for sparse GPs, i.e. $T^{NN}_{\dot{v}_y} = 30$ and $T^{NN}_{\ddot{\theta}} = 50$ for $\dot{v}_y$ and $\ddot{\theta}$, respectively, allows the controller to run at a control frequency $f_c = 20Hz$ on the real go-kart.

\review{In such complex experimental scenarios,} theoretical guarantees on the stability of the algorithm are hardly achievable. \review{Moreover, the adoption of a black-box model identified in the continuous domain and then discretized further complicates the mathematical description. However,} our approach \textit{empirically} supports the recursive feasibility of NLP by adopting the following expedients:
\begin{enumerate}
    \item a soft constraint on $e_y$, which is the only constraint where the model uncertainty is emphasized, that, by marginally reducing the bound value, allows for a collision-free trajectory;
    \item a sufficiently long prediction horizon that includes the next curve, allowing the controller to compute a velocity profile compatible with the maximum lateral acceleration and the track boundaries;
    \item a final reference with high weights on lateral and angular errors $e_y$ and $e_\psi$, exploiting the well-known effect of the final cost to lead to conservative behavior.
\end{enumerate}

\begin{figure}[t]
    \centering
    \includegraphics[width=0.85\linewidth]{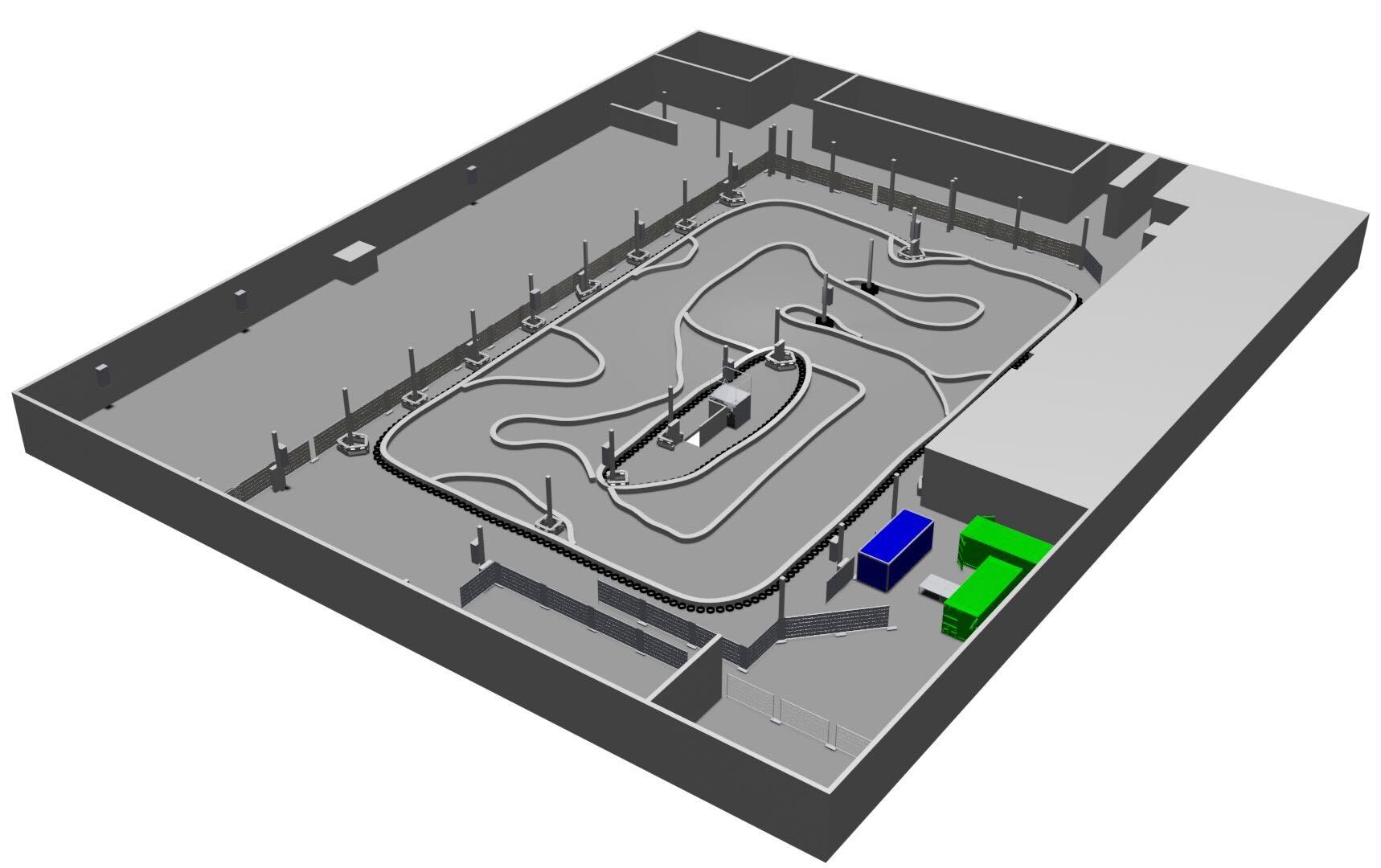}
    \caption{The simulation environment in Gazebo.}
    \label{fig:simulation_environment}
\end{figure}
   
    \begin{figure}[t]
    \centering
    \includegraphics[width=\linewidth]{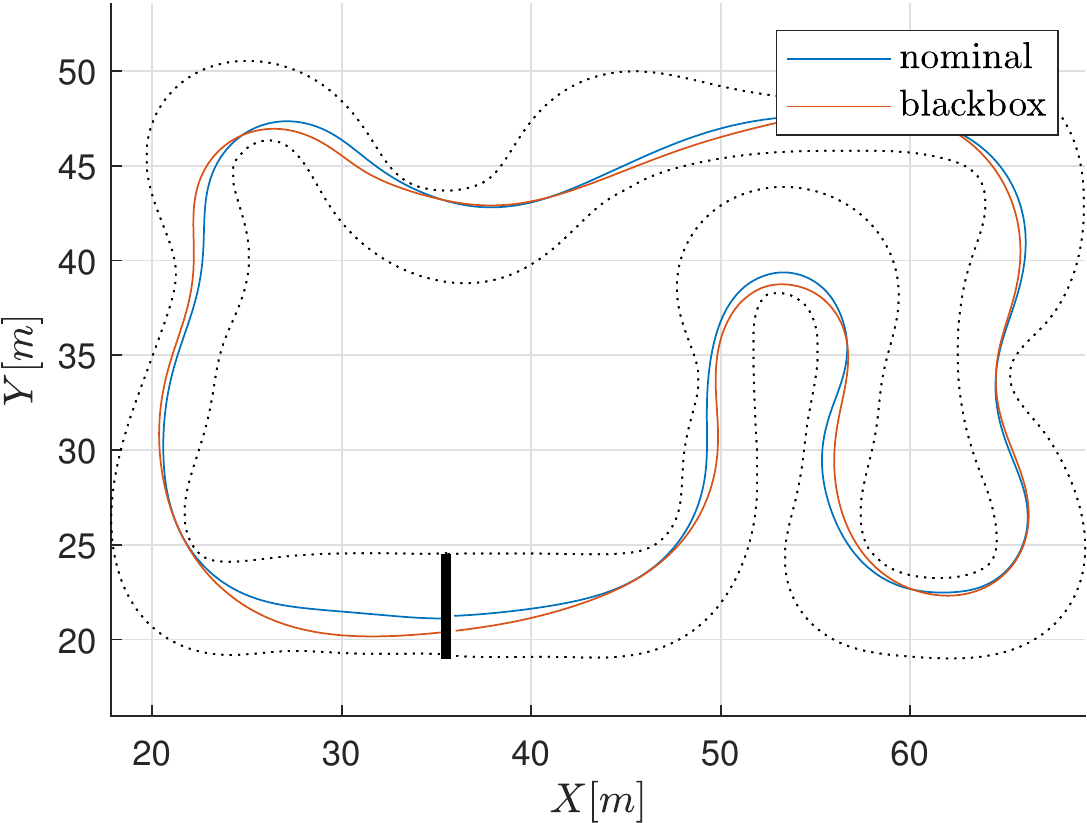}
    \caption{Comparison of trajectory obtained by the (Lb)NMPC using the nominal and black-box models in simulation.}
    \label{fig:sim_cmp_XY}
    \end{figure}
    
    \begin{figure}[t]
    \centering
    \includegraphics[width=\linewidth]{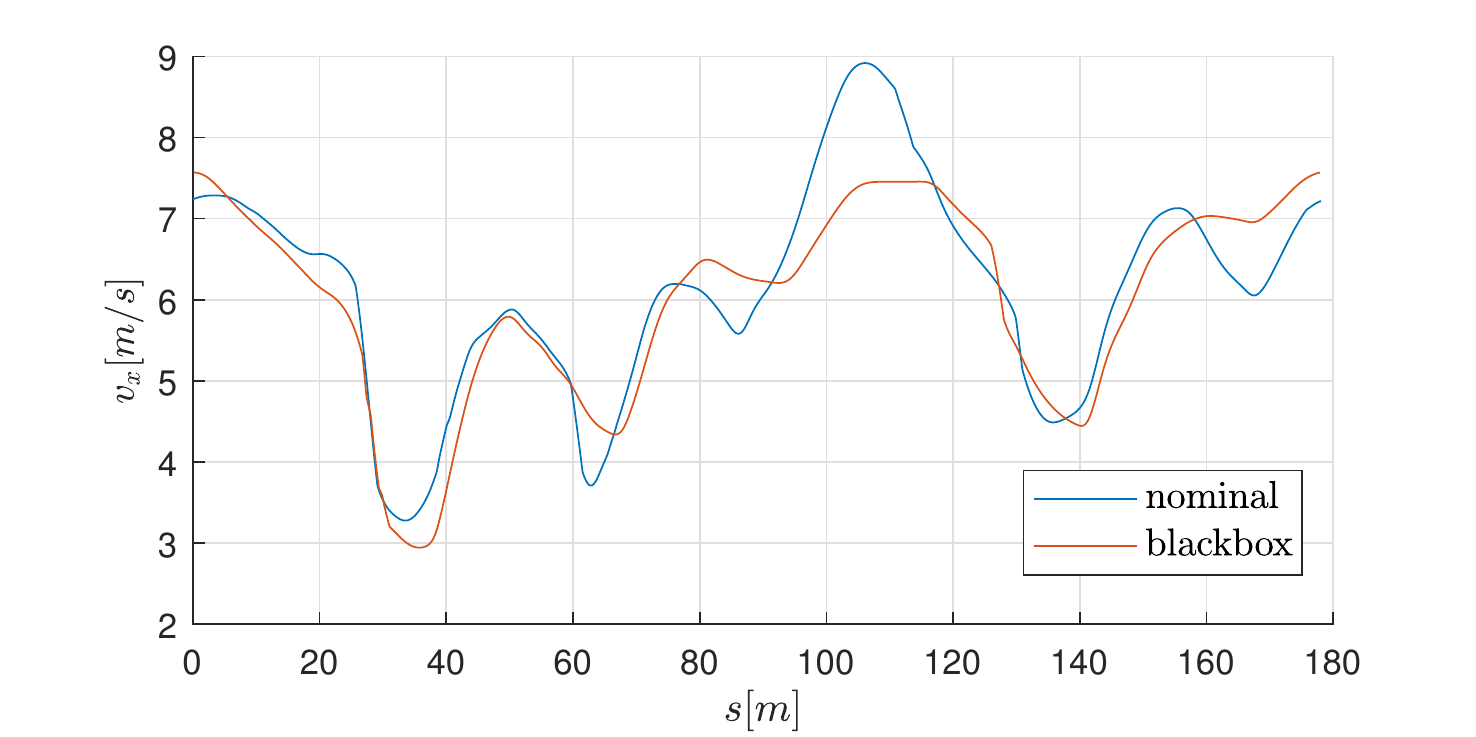}
    \caption{Comparison of longitudinal velocity obtained by the (Lb)NMPC using the nominal and black-box models in simulation.}
    \label{fig:sim_cmp_vx}
    \end{figure}

\subsection{Preliminary simulation study}\label{sec:sim}
A preliminary study in simulation has been accomplished in order to validate the feasibility of the developed strategy and identify a tuning set of the hyperparameters, to be used as a starting point for the experimental scenario in a completely controllable environment. In particular, the control algorithm is forced to fulfill real-time requirements while no sensor noise and actuation delays are present, allowing setting the GP number of point $T^{NN}$, the prediction horizon length $N$, the integration step $T_s$ and the grid $G$.

A previously developed simulation environment \cite{rohit}, \cite{ankenbrand}, based on \textit{C++} and Gazebo within \textit{ROS} (Robot Operating System) \cite{quigley2009ros}, has been used for this phase (see Fig. \ref{fig:simulation_environment}). It uses the same communication protocol and reproduces the same channels as the real go-kart, allowing for a practical test of the controller as it would be in the experimental scenario. 

\begin{table}[b]
    \centering
    \caption{RMSE of online one-step-ahead velocity predictions $\hat{e}_{\dot{q}}$ for nominal and black-box model in simulation.}
    \label{tab:RMSE_simulation}
    \begin{tabular}{lcc}
        model & $v_y$ $[m/s]$ & $\dot{\theta}$ $[rad/s]$  \\
        \hline\\
        nominal & $4.03 \cdot 10^{-2}$ & $4.89 \cdot 10^{-2}$ \\
        black-box & $2.04 \cdot 10^{-2}$ & $4.57 \cdot 10^{-2}$
    \end{tabular}
\end{table}

The LbNMPC controller has been run in the simulation framework, adopting the same configuration that will be tested on the real go-kart, except for the black-box prediction model within. 
As expected, the black-box model is fairly precise in predicting online the one-step-ahead velocities for both $v_y$ and $\dot{\theta}$, as shown in Tab. \ref{tab:RMSE_simulation}.  In particular, the error is in the same order of magnitude as the nominal model, counting half the error for $v_y$ and slightly lower for $\dot{\theta}$, confirming that the GPs are actually describing the system dynamics, in average, accurately.
This fact is reflected in a superior closed-loop behavior of the black-box strategy in some curves, but yet worse in others. Indeed, the driven paths are quite similar (see Fig. \ref{fig:sim_cmp_XY}), while the turns are traveled at different velocities (shown in Fig. \ref{fig:sim_cmp_vx}), leading to nearly the same lap-time overall, i.e. $26.80s$ for black-box model and $26.55s$ for the nominal model.
The obtained behaviour is relevant in the perspective of using the scheme in the experimental environment since it illustrates the validity of the proposed approach.

\subsection{Experimental Results}\label{subsec:exp}

\begin{figure}[tb]
    \centering
    \vspace{0.1cm}
    \includegraphics[width=0.9\linewidth,trim=0 0 0.1cm 0, clip]{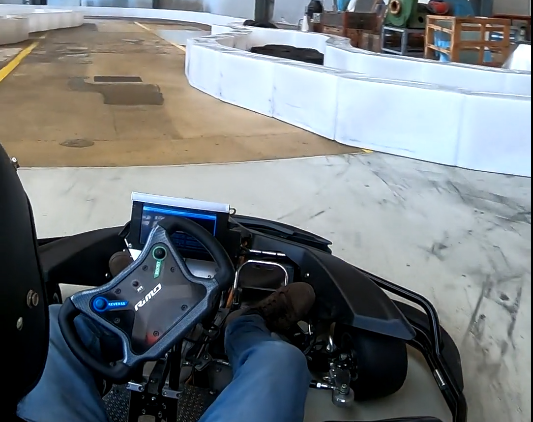}
    \caption{The real go-kart platform driving autonomously.}
    \label{fig:gokart_driving}
\end{figure}

The LbNMPC controller based on the model presented in Sec. \ref{subsubsec:model_exp} has been implemented and tested on the real go-kart platform. An indoor 180m long track has been used as a test bench, comparing the performance obtained using the black-box modeling with respect to the nominal ones. 
In the experimental scenario relevant approximations for the data-driven modeling have been observed, therefore reducing prediction capabilities. 
In particular, the higher complexity of the learned functions leads to higher errors when applying the NN local approximation (as reported in Tab. \ref{tab:comp_err}).
The mean time for solving the LbNMPC problem resulted in $28.53ms$, while it resulted in $10.21ms$ for the nominal NMPC one. 

\subsubsection{Go-kart Platform}
the go-kart is based on a RiMO SiNUS iON electric rear wheel-driven go-kart platform. The vehicle is equipped with different sensors for state estimation and localization such as IMU and LiDAR. It mounts a custom computer on the back, i.e. an Intel Xeon D-1540 @2.00Ghz CPU with 8 cores. The computer interfaces with the sensors and the actuators, then it processes the data and runs the localization, mapping, and state estimation modules while executing the controller. The communication between the different frameworks is done using ROS which gives a structured communication layer above the Ubuntu operating system.

\subsubsection{Results on Track}

\begin{figure}
    \centering
    \includegraphics[width=.9\linewidth]{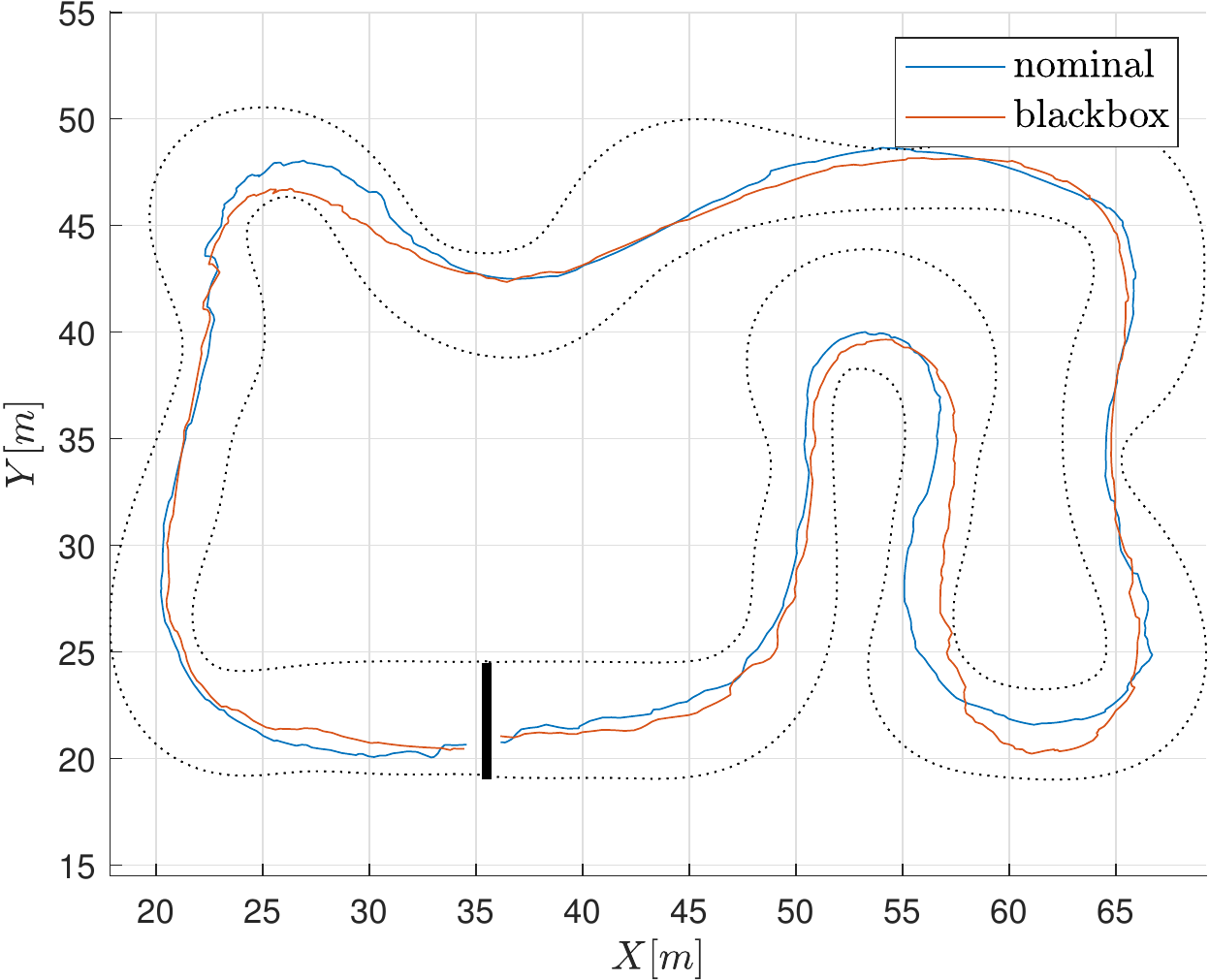}
    \caption{Comparison of trajectory obtained by the (Lb)NMPC using the nominal and black-box models on the real go-kart.}
    \label{fig:cmp_XY}
\end{figure}
\begin{figure}
    \centering
    \includegraphics[width=.9\linewidth]{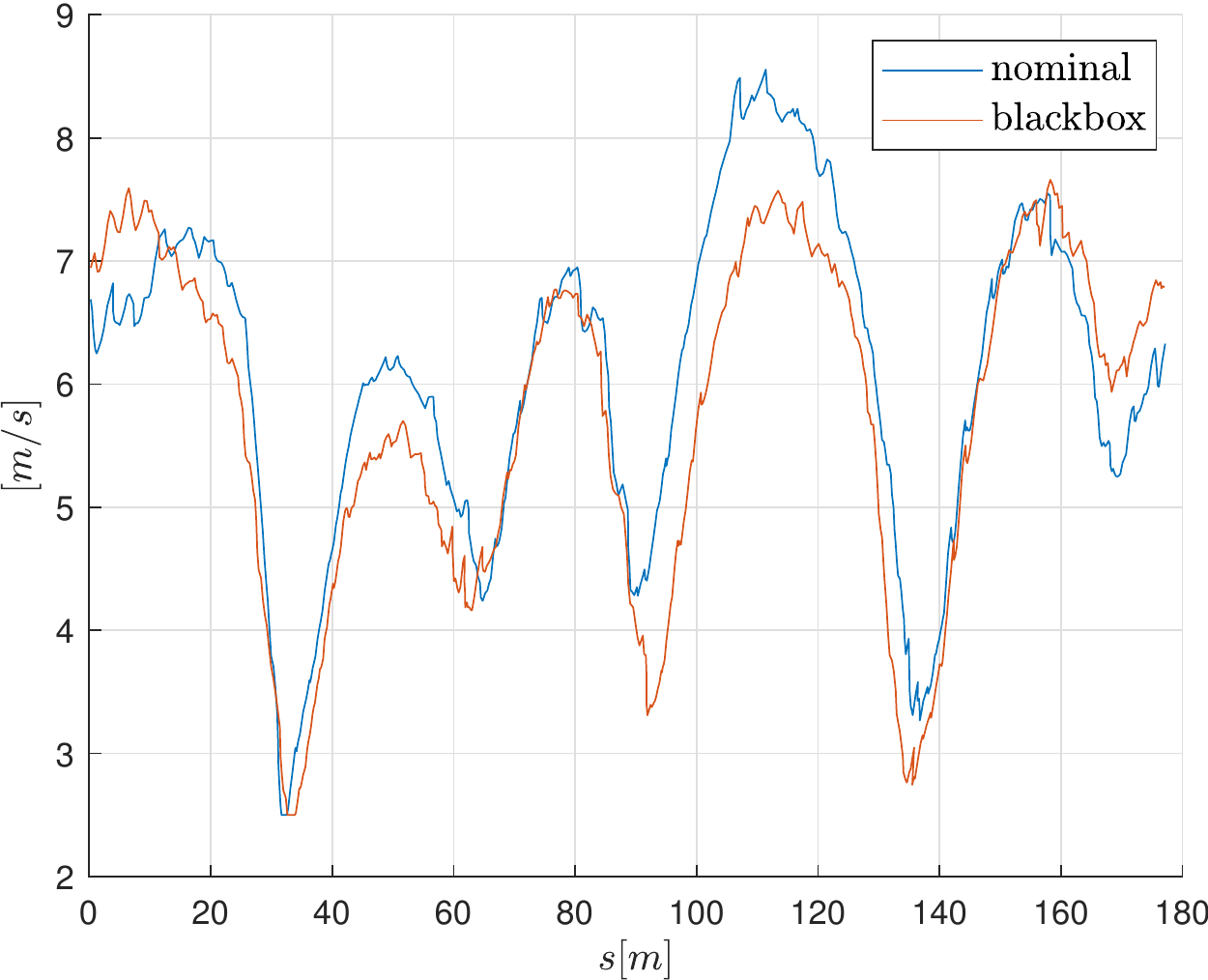}
    \caption{Comparison of longitudinal velocity obtained by the (Lb)NMPC using the nominal and black-box models on the real go-kart.}
    \label{fig:cmp_vx}
\end{figure}
\begin{figure}
    \centering
    \includegraphics[width=.9\linewidth]{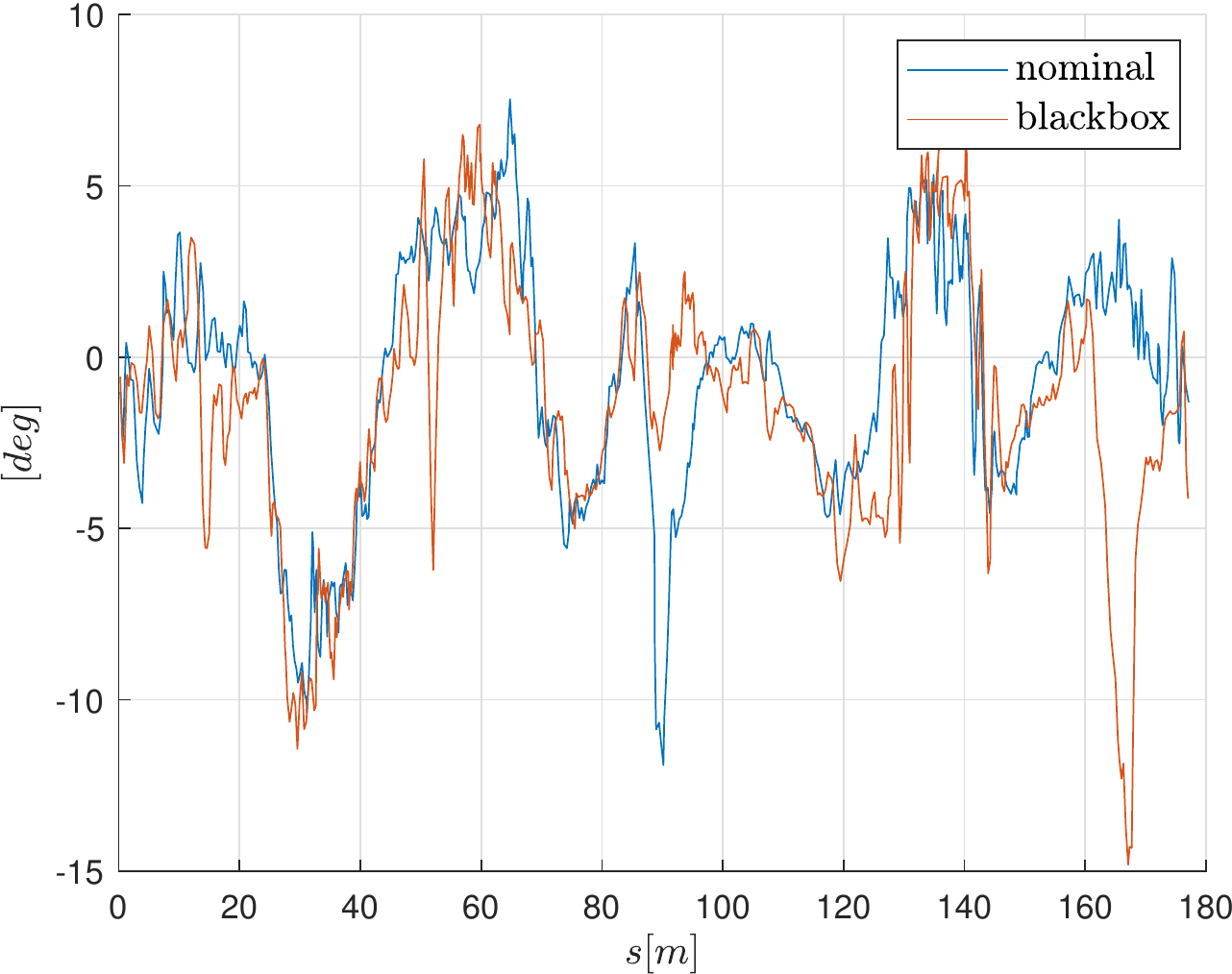}
    \caption{Comparison of sideslip obtained by the (Lb)NMPC using the nominal and black-box models on the real go-kart.}
    \label{fig:cmp_ss}
\end{figure}
\begin{figure}
    \centering
    \includegraphics[width=.9\linewidth]{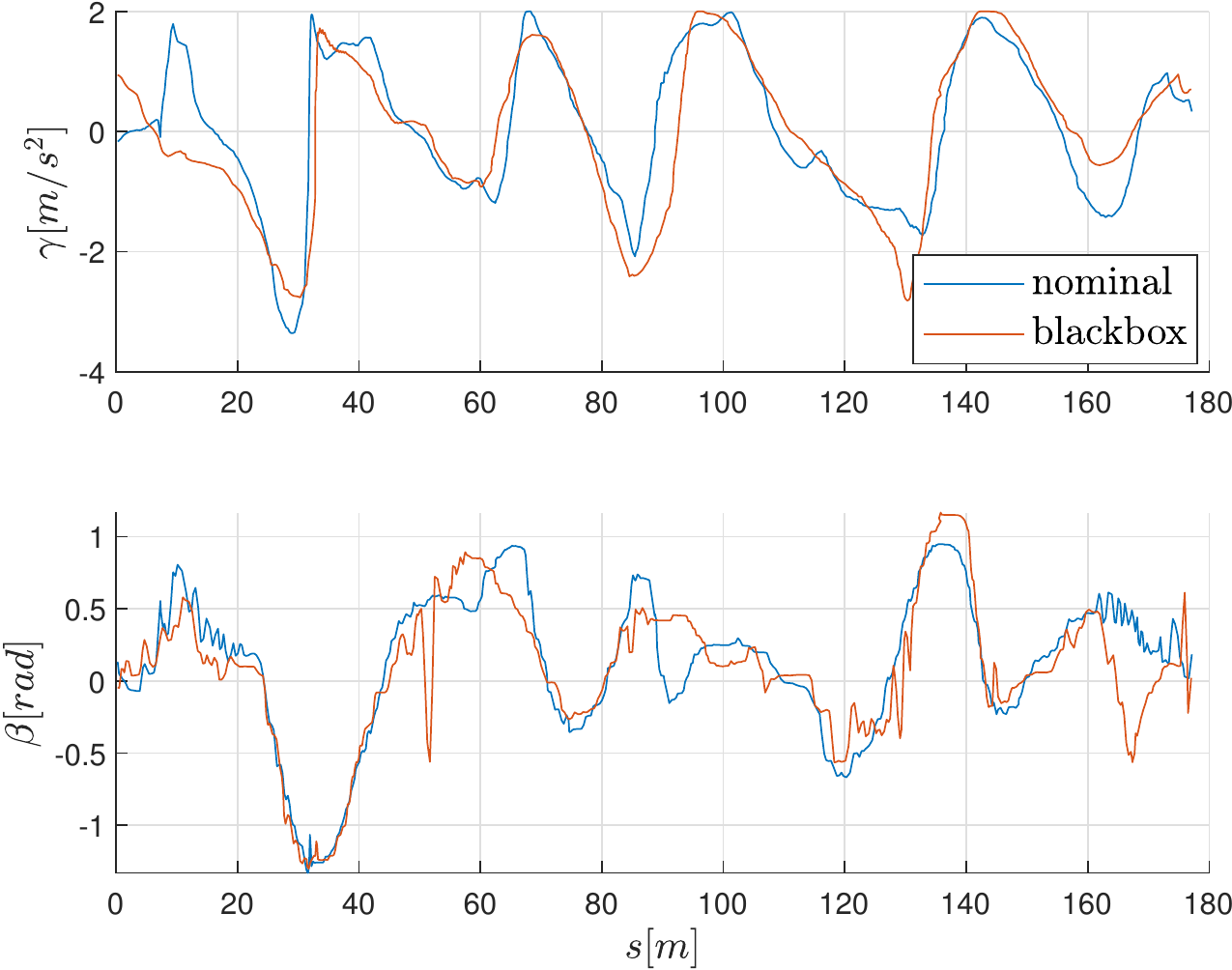}
    \caption{Comparison of controls computed by the (Lb)NMPC using the nominal and black-box models on the real go-kart.}
    \label{fig:cmp_ctrl}
\end{figure}
\begin{figure}
    \centering
    \includegraphics[width=.9\linewidth]{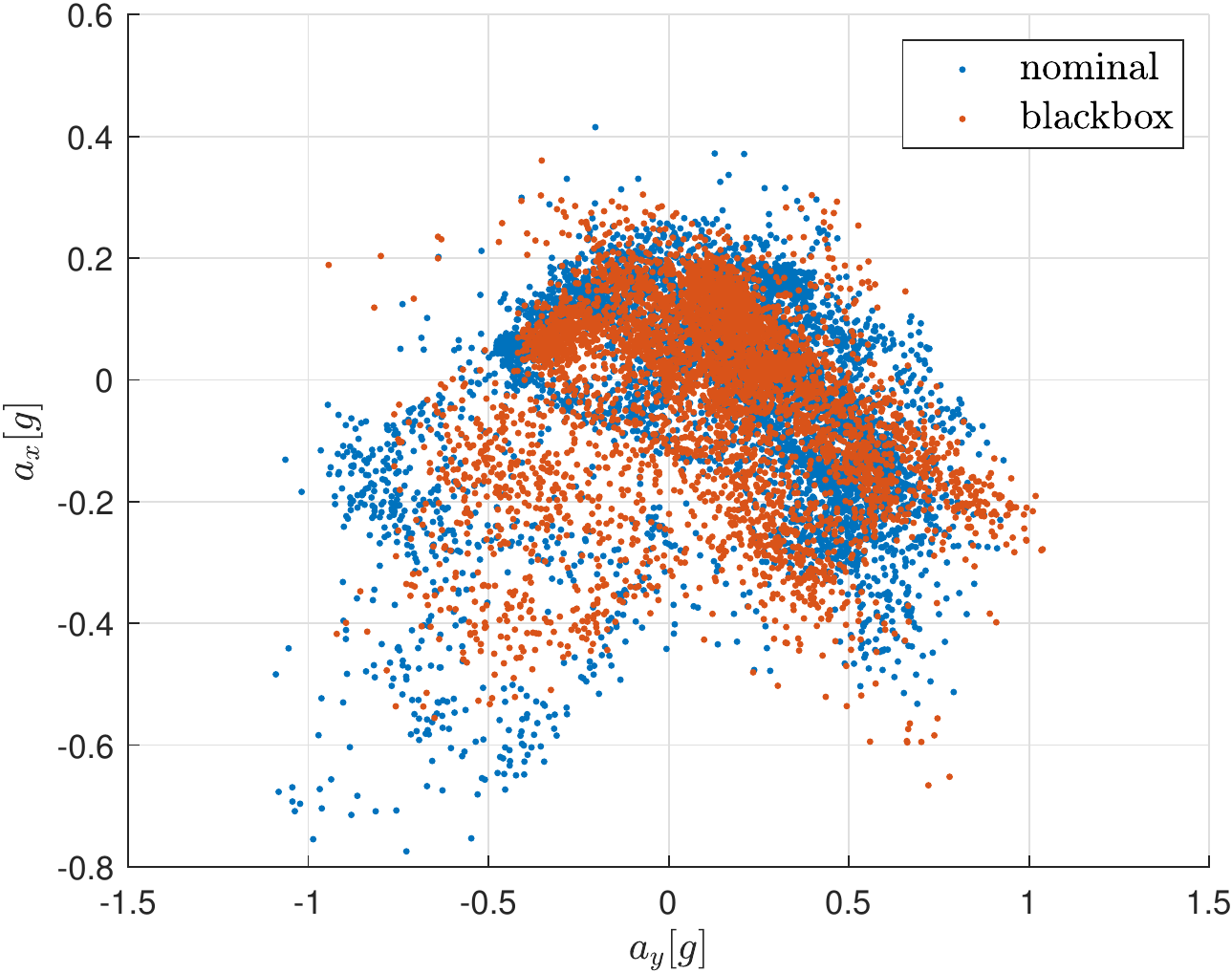}
    \caption{Comparison of gg-diagram (longitudinal vs lateral accelerations) obtained by the (Lb)NMPC using the nominal and black-box models on the real go-kart.}
    \label{fig:cmp_gg}
\end{figure}

\begin{table}[b]
    \centering
    \caption{RMSE of online one-step-ahead velocity predictions $\hat{e}_{\dot{q}}$ for nominal and black-box model on real go-kart.}
    \label{tab:RMSE_exp}
    \begin{tabular}{lcc}
        model & $v_y$ $[m/s]$ & $\dot{\theta}$ $[rad/s]$  \\
        \hline\\
        nominal & $7.63 \cdot 10^{-2}$  & $5.97 \cdot 10^{-2}$ \\
        black-box & $9.80 \cdot 10^{-2}$ & $8.05 \cdot 10^{-2}$
    \end{tabular}
\end{table}

the LbNMPC presented in the previous section has been employed to drive the go-kart along the track, depicted in Fig. \ref{fig:cmp_XY}, obtaining comparable results with respect to the nominal strategy. Specifically, the driven trajectories resulted very similarly over most of the track, 
with slight differences in specific parts. 
The path, velocity, sideslip and controls are shown in Fig. \ref{fig:cmp_XY}-\ref{fig:cmp_ctrl}.
In particular, the black-box approach travels the bottom-left turn (at $160-180m$) with higher velocity, controlling a high sideslip, gaining speed over the whole bottom straight.
The two schemes obtain similar performance at the top-left corner ($120-150m$) and between the first and the second U turn (at $40-60m$), with different strategies. In both cases, the black-box controller is cutting the curve more than the nominal, travelling less distance but slightly lowering the speed at the curve center. 
On the other hand, in the top-right corner (at $80-100m$) the center of the turn is travelled at a lower velocity by the black box strategy, making the whole top straight with a speed gap with respect to the nominal controller.
This behaviour results in a lap time of $29.3s$ for black-box, that is $3.75\%$ slower than the nominal lap time, i.e. $28.2s$. Interestingly, both models resulted in similar behavior under sideslip conditions (Fig. \ref{fig:cmp_ss}), with the black-box model allowing to stabilize the vehicle after an oversteer with a side slip of $15deg$ during the exit of the bottom-left curve, at $160-180m$.  The actual commands sent to the go-kart using the different modeling, shown in Fig. \ref{fig:cmp_ctrl}, are mostly consistent, except for the counter-steering actions at $50m$ and $170m$ for the black-box controller and at $90m$ for the nominal one. 
The maximum accelerations are similar for both control strategies (Fig. \ref{fig:cmp_gg}), with peaks of $\pm 1g$ for lateral and $[+0.2,-0.6]g$ for longitudinal (i.e. the longitudinal maximum accelerations given by motor and braking system) respectively.
In this framework, the nominal model is more precise on average in predicting velocities with respect to the black-box one, i.e. velocity errors around $25\%$ lower, as reported in table \ref{tab:RMSE_exp}: indeed, the GP approximations needed to obtain a real-time controller led to a relevant reduction in the prediction capabilities. However, the obtained system representation is sufficiently informative to effectively complete the driving task on the experimental go-kart. It is expected that, it could be further improved by providing more computing resources.
\begin{figure}
    \centering
    \includegraphics[width=.9\linewidth]{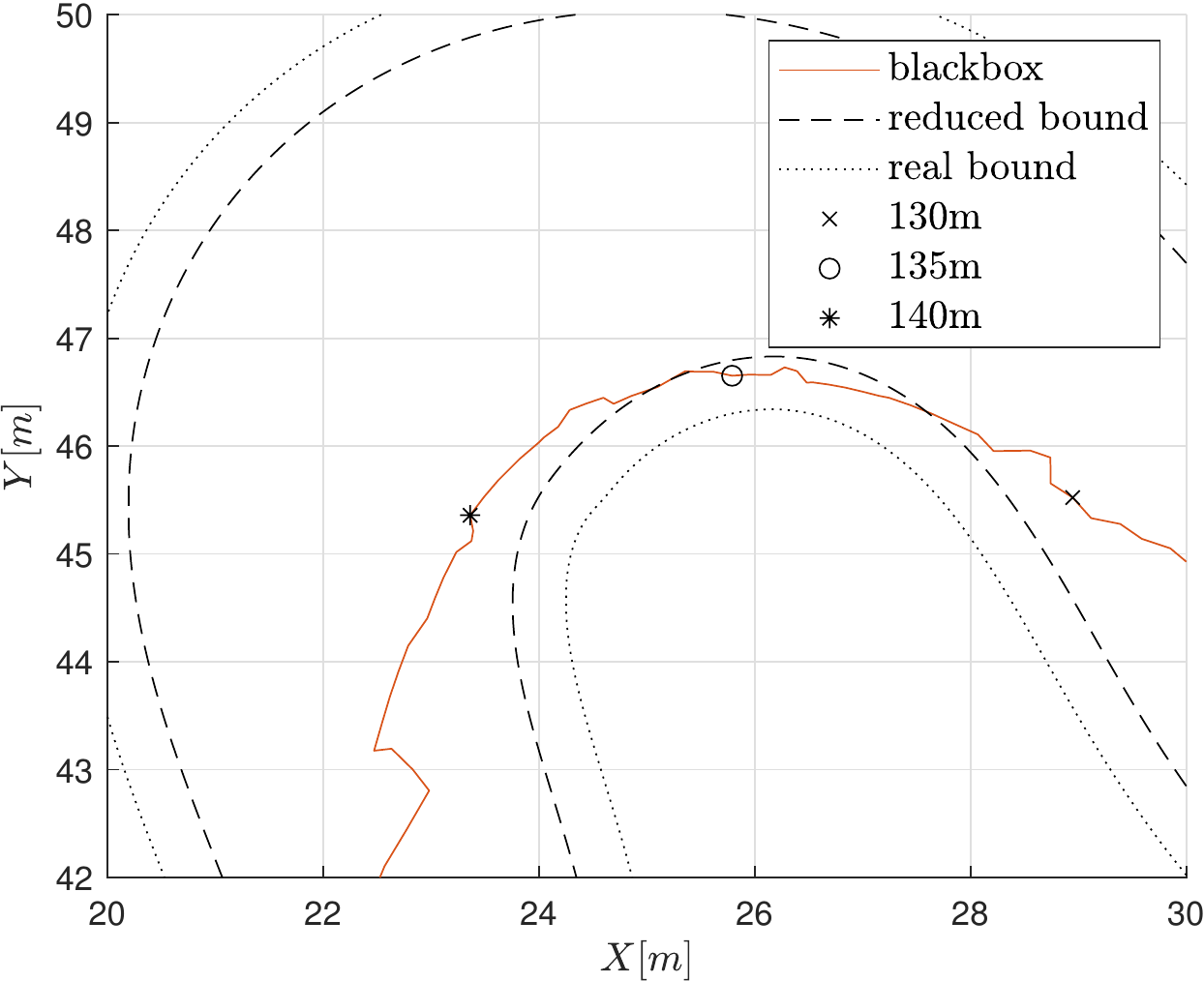}
    \caption{Detail of the trajectory travelled by the LbNMPC controller, highlighting the reduced bound on lateral error during the top left curve.}
    \label{fig:feas_XY}
\end{figure}
\begin{figure}
    \centering
    \includegraphics[width=.9\linewidth]{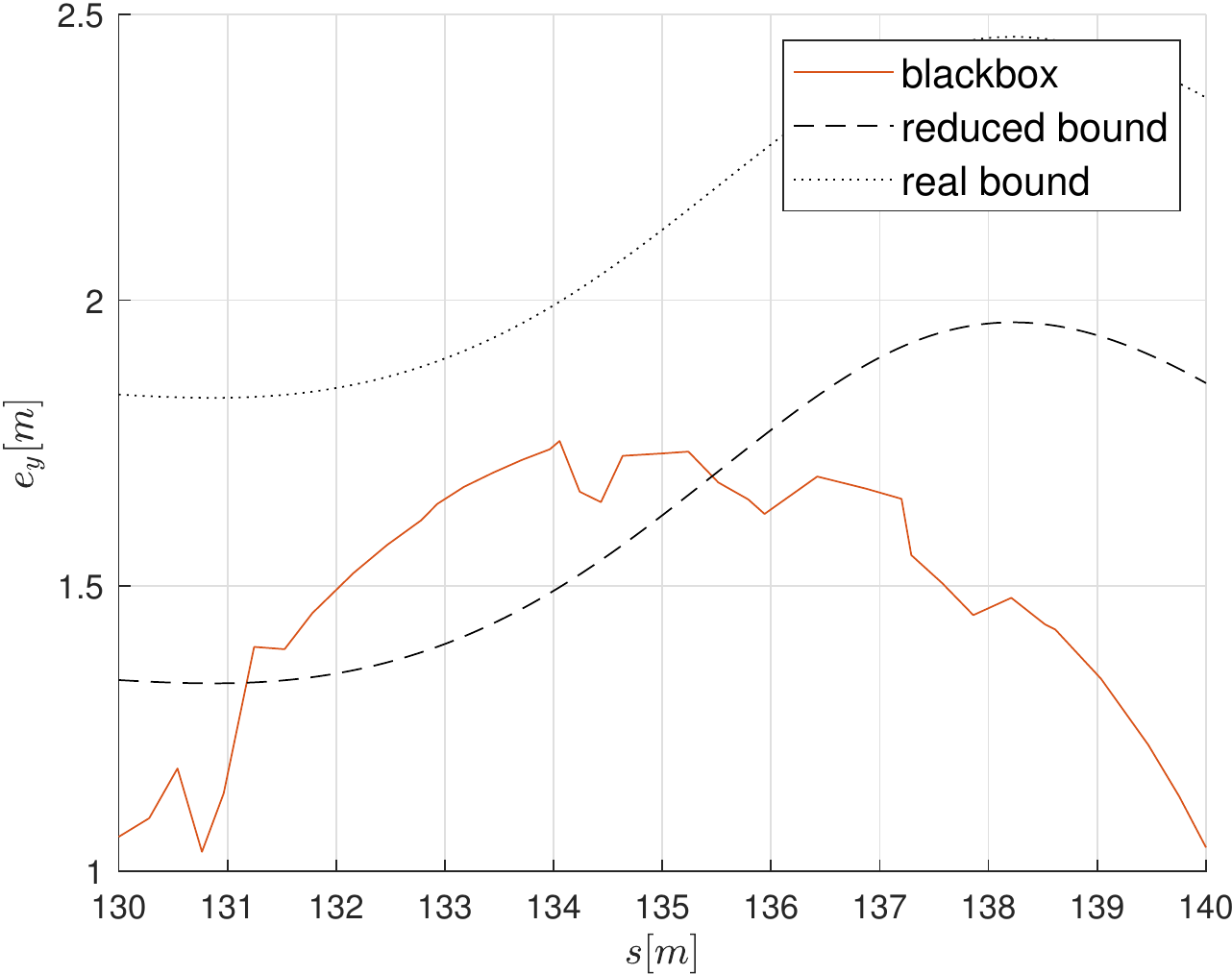}
    \caption{Detail of the lateral error value and its reduced bound during the top left curve.}
    \label{fig:feas_ey}
\end{figure}
\begin{figure}
    \centering
    \includegraphics[width=.9\linewidth]{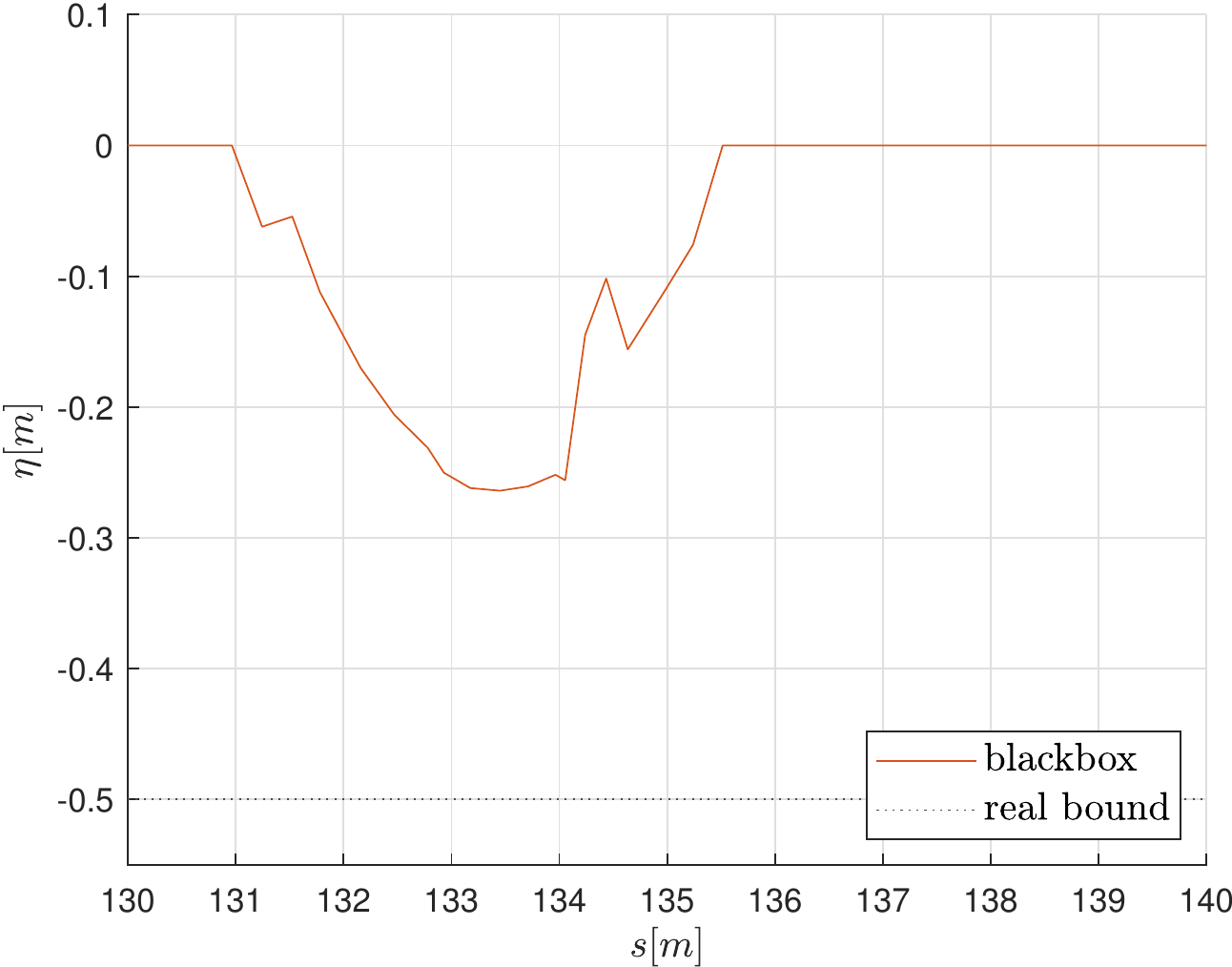}
    \caption{Detail of the slack variable value during the top left curve.}
    \label{fig:feas_slack}
\end{figure}

\subsubsection{Empirical Feasibility Validation}
the expedients employed to empirically achieve recursive feasibility, i.e. avoiding hitting track boundaries, of the NLP described in Sec. \ref{subsec:ctrl_setup} have been validated to establish their effectiveness. In particular, the effect of soft constraint on the lateral error can be clearly observed in the top-left turn of the track. In Fig. \ref{fig:feas_XY}-\ref{fig:feas_slack} details of trajectory, lateral error, and slack variable values are shown.
While travelling the curve, the go-kart is very close to the left bound of the track, and the slack variable is needed to ensure the feasibility of the problem, while, at the same time, pushing the controller to move away from the limit. In fact, once the \textit{reduced bound} has been overcome, the slack variable is activated, avoiding both the controller to fail and the go-kart to crash.

\section{Conclusion}\label{sec:conclusion}

A LbNMPC controller for a go-kart based on a black-box model of the accelerations obtained by Gaussian Processes has been presented. The formulation exploits IMU and localization data collected while a human is driving to obtain GP models of the dynamics, eliminating the need for an a priori known dynamics model. The computational burden of this method was addressed by reducing the GP size, exploiting local approximations, and applying fast NMPC solutions.
The strategy achieved satisfactory results on a 180m long indoor track, comparable to a NMPC based on a detailed nominal dynamic model, although with slightly lower performances, limited by the needed approximations. However, the experiment demonstrates the capability of this approach to effectively control a complex system in a real-world scenario. In particular, the possibility of using a black-box approach for modeling a 4-wheel vehicle for use in an NMPC controller was demonstrated.
Further development will include an analysis of different \textit{offline} reduction methods and \textit{online} local approximations for GP, beyond the usage of \textit{adaptive} models leveraging the data acquired in real-time.

\bibliographystyle{IEEEtran}
\bibliography{IEEEabrv,tcst22}

\end{document}